%% file: Note2D.tex
\documentclass[11pt,a4paper]{article}

\usepackage[a4paper, hmargin={2.2cm,2.2cm}, vmargin={2.2cm,2.2cm}]{geometry}
\usepackage{amsmath,amsfonts,amssymb}

\usepackage{graphicx}

\usepackage[style=phys, sorting=none, backend=biber]{biblatex}
\bibliography{references}

\usepackage{float}
\usepackage{pgfplots}
\pgfplotsset{width=10cm,compat=1.9}
\usepgfplotslibrary{groupplots}

\makeatletter
\@fpsep\textheight
\makeatother

\tikzstyle{style_1}=[mark = triangle, only marks]
\tikzstyle{style_2}=[mark = square, only marks]
\tikzstyle{style_3}=[mark = diamond, only marks]
\tikzstyle{style_4}=[mark = pentagon, only marks]
\tikzstyle{style_5}=[mark = o, only marks]
\tikzstyle{style_6}=[mark = oplus, only marks]
\tikzstyle{style_7}=[mark = otimes, only marks]

\pgfplotscreateplotcyclelist{my BandW}{%
solid, every mark/.append style={solid, fill=gray}, mark=triangle*\\%
solid, every mark/.append style={solid, fill=gray}, mark=square*\\%
solid, every mark/.append style={solid, fill=gray}, mark=diamond*\\%
solid, every mark/.append style={solid, fill=gray}, mark=pentagon*\\%
solid, every mark/.append style={solid, fill=gray}, mark=*\\%
solid, every mark/.append style={solid, fill=gray}, mark=oplus*\\%
solid, every mark/.append style={solid, fill=gray}, mark=otimes*\\%
solid, every mark/.append style={solid, fill=gray}, mark=halfcircle*\\%
dotted, every mark/.append style={solid, fill=gray}, mark=square*\\%
densely dotted, every mark/.append style={solid, fill=gray}, mark=otimes*\\%
loosely dotted, every mark/.append style={solid, fill=gray}, mark=triangle*\\%
dashed, every mark/.append style={solid, fill=gray},mark=diamond*\\%
loosely dashed, every mark/.append style={solid, fill=gray},mark=*\\%
densely dashed, every mark/.append style={solid, fill=gray},mark=square*\\%
dashdotted, every mark/.append style={solid, fill=gray},mark=otimes*\\%
dashdotdotted, every mark/.append style={solid},mark=star\\%
densely dashdotted,every mark/.append style={solid, fill=gray},mark=diamond*\\%
}

\usepackage{booktabs}
\newcommand{\ra}[1]{\renewcommand{\arraystretch}{#1}}

\usepackage{hyperref}
\hypersetup{
    colorlinks=false,
    linkcolor=blue,
    filecolor=magenta,      
    urlcolor=cyan,
}

\newcommand{\quoting}[1]{``#1''}
\newcommand{\td}{\text{d}}

\usepackage{xspace}
\newcommand{\zeroth}{zeroth\xspace}
\newcommand{\first}{first\xspace}
\newcommand{\second}{second\xspace}

\let\oldsqrt\sqrt
\def\sqrt{\mathpalette\DHLhksqrt}
\def\DHLhksqrt#1#2{%
\setbox0=\hbox{$#1\oldsqrt{#2\,}$}\dimen0=\ht0
\advance\dimen0-0.2\ht0
\setbox2=\hbox{\vrule height\ht0 depth -\dimen0}%
{\box0\lower0.4pt\box2}}

\begin{document}

\hypersetup{pageanchor=false}
\begin{titlepage}
\begin{center}
\quad\par
\vspace{4cm}
{\LARGE\bfseries Consistency in Drift-ordered Fluid Equations\par}
\vspace{1cm}
{\bfseries Jakob Gath and Matthias Wiesenberger\par}
\vspace{1cm}
{Department of Physics, Technical University of Denmark, \\ DK-2800 Kgs. Lyngby, Denmark\par}
\vspace{1cm}
{\href{mailto:jagath@fysik.dtu.dk}{jagath@fysik.dtu.dk}, 
\href{mailto:mattwi@fysik.dtu.dk}{mattwi@fysik.dtu.dk}\par}
\vspace{2cm}
{\bfseries \large Abstract\par}
\vspace{0.4cm}
\end{center}

\noindent We address several concerns related to the derivation of drift-ordered fluid equations. Starting from a fully Galilean invariant fluid system, we show how consistent sets of perturbative drift-fluid equations in the case of a isothermal collisionless fluid can be obtained. Treating all the dynamical fields on equal footing in the singular-drift expansion, we show under what conditions a set of perturbative equations can have a non-trivial quasi-neutral limit. We give a suitable perturbative setup where we provide the full set of perturbative equations for obtaining the \first-order corrected fields and show that all the constants of motion are preserved at each order. With the dynamical field variables under perturbative control, we subsequently provide a quantitative analysis by means of numerical simulations. With direct access to first-order corrections the convergence properties are addressed for different regimes of parameter space and the validity of the \first-order approximation is discussed in the three settings: cold ions, hot ions and finite charge density.
	
\end{titlepage}
\hypersetup{pageanchor=true}

\newpage
\tableofcontents
\noindent\hrulefill

\vspace{0.2cm}

\section{Introduction}

The scrape-off layer in magnetically confined plasmas is typically treated as a cold and collision-rich region where the turbulence is characteristically low frequency and long wavelength in nature. In this regime, fluid models are arguably widely applicable \cite{Braginskii, doi:10.1063/1.1693260, doi:10.1063/1.872368, doi:10.1063/1.4943199}, as they are suitable for studying the stability and transport properties of the plasma on scales much longer than the ion gyroradius and much lower than the cyclotron frequency. In fact, many key characteristics of the scrape-off layer are captured by the drift-fluid theory (see e.g. \cite{PhysRevLett.81.4396}). For drift-fluid models, it is common to treat the fluid-momentum equation in a singular perturbation expansion that makes it possible to solve explicitly for the perpendicular components of the fluid velocity. Together with the additional fluid conservation equations this constitutes the so-called \textit{drift-ordered} fluid equations. Perturbative fluid equations derived in this way are used in many plasma physics simulation codes \cite{TAMAIN2010361, GBScode, RasmussenHESEL}. However, there are several fundamental concerns with respect to the consistency of such perturbative schemes:

\vspace{0.2cm}

\textit{(i)} In the standard derivation of drift-ordered fluid equations, the dynamical field variables are (questionably) rarely expanded on equal footing. It is therefore impossible to systematically check how well the approximation represents the true underlying system as there is no notion of \quoting{order}. To control the expansion in a coupled non-linear system, all \textit{dynamical} fields should in general be treated on equal footing such that the corrections to the dynamics appear in an ordered scheme. In this way, one can be monitor the corrections and estimate whether a given order of approximation has converged to a sufficient degree.

\textit{(ii)} A ionized plasma is primarily governed by the electromagnetic force. In many cases, one is therefore interested in systems where (at least a part of) the electromagnetic fields are self-consistently obtained alongside the non-relativistic fluid variables, e.g. the electric field sustained by a plasma in a background magnetic field. However, electromagnetism is a Lorentz invariant theory and therefore a suitable non-relativistic limit of the Maxwell equations is required to obtain a coupled system (with the non-relativistic fluid equations) that is fully invariant under Galilean transformations.

\textit{(iii)} One of the defining criteria for an ionized gas to be characterized as a plasma is to have a sufficiently high density such that local concentrations of charge is shielded out at very short distances. In such cases, the plasma can be said to exist in a quasi-neutral state, where the densities of the charged species are equilibrated. However, all interesting electromagnetic phenomena ultimately relies on separation of charge. There exists therefore a subtle interplay between the quasi-neutral property and the dynamics of the plasma. For this reason, it is important that the quasi-neutral assumption is implemented on the same footing along with any other assumption such that it becomes a natural consequence of the perturbative setup and comes under perturbative control. In general, if an assumption is implemented before treating the full underlying system of equations, the perturbative nature of the assumption is lost. 

\textit{(iv)} Treating the fluid-momentum equation as a singular perturbation problem allows for a separation between the fast and slow dynamics and the construction of the drift expansion. However, since the momentum equation is now degenerate it can not possible satisfy all initial conditions that can be specified for the full underlying fluid system (i.e. the solution space is naturally restricted). The initial conditions will instead be completely determined by the initial conditions for the other dynamical fields. One should therefore make sure that the perturbation problem under consideration is not ill-posed.

\vspace{0.2cm}

Before proceeding, we start with a general discussion about the quasi-neutral assumption in drift-ordered fluid models. In the general drift-ordered setup, the perpendicular components of the velocity are decomposed into three primary terms; the $\text{E}{\times}\text{B}$, the diamagnetic, and the polarization-drift velocity. The latter two manifest the separation of charge in the system. However, under a strict quasi-neutral assumption no charge separation in the system is in principle allowed which gives rise to some interesting constraints on the possible ordering of the drift velocities to ensure the consistency of the system of equations. Typically, the $\text{E}{\times}\text{B}$-drift velocity is taken to be the main contribution to the velocity field (and advection of density) while the polarization-drift velocity by construction of the singular perturbation expansion must appear at higher order. Now, in many drift-fluid models the diamagnetic-drift velocity is typically assumed to be of the same magnitude as the $\text{E}{\times}\text{B}$-drift velocity. However, in a proper perturbative setup this would require the charge separation to be treated at the same order as the $\text{E}{\times}\text{B}$-drift velocity which is in direct contradiction with the assumption that the separation of charge is negligible. Such cases can only make sense for sufficiently dilute plasmas where the quasi-neutral assumption can be lifted (or if the diamagnetic-drift velocity does no convection as in a constant magnetic field configuration \cite{doi:10.1063/1.1693260}). Otherwise, it will have to appear at the succeeding order where it in conjunction with the polarization-drift velocity provides a time evolution for the electric potential. In fact, the quasi-neutral assumption reduces the charge-current conservation equation to the divergenceless current constraint and only due to the polarization-drift velocity this appears as a dynamical constraint equation. Alternatively, it is natural to attempt to push the convection due to the diamagnetic-drift velocity to the next order as it is arguably small (in a nearly constant magnetic field). However, it turns out that this will push the polarization-drift velocity to an even higher order reducing the current conservation equation to a purely spatial constraint that can only be satisfied by a constant magnetic field or zero pressure gradients.

\vspace{0.2cm}

We go through the steps of obtaining a consistent perturbative expansion of a simple collisionless isothermal fluid governed by the Lorentz force in Sec.~\ref{sec:forcedfluid} that address all of the above listed points. Subsequently, we find solutions to the perturbative system using the numerical tools available in the Feltor library \cite{feltor} in the case of a two species fluid in Sec.~\ref{sec:simulation}. Through the numerical solutions we perform a quntitative analysis and estimate the physical regime of validity of the presented perturbative system. We include three appendices. In App.~\ref{sec:nonrel}, we show how a consistent non-relativistic limit of the Maxwell equations can be obtained. In App.~\ref{sec:charges}, we provide some details on the conserved charges of the forced fluid system presented in Sec.~\ref{sec:forcedfluid}. In App.~\ref{sec:alternativeperb}, two alternative perturbative setups valid for dilute plasmas (in contrast with the equations given in Sec.~\ref{sec:scheme2}) are given.

\section{Non-relativistic fluid equations} \label{sec:forcedfluid}

A charged non-relativistic fluid in (3+1)-dimensions is constituted by the continuity, momentum, energy and and charge-current conservation equations. Here, we will consider a collisionless \textit{isothermal} fluid system without dissipative terms of charged particles governed by the Lorentz force,
\begin{align} \label{eq:fluidsystem}
\begin{split}
\partial_t n_a + \nabla \cdot \left( n_a \mathbf{v}_a \right) = 0 ~, \quad
m_a \td_t \mathbf{v}_a = -\bar{T}_a \nabla \log(n_a) + q_a \left( -\nabla \phi + \mathbf{v}_a \times \mathbf{B} \right) ~.
\end{split}
\end{align}
The latin subscript $a$ labels the species of the particle and bold symbols are spatial vectors. The total derivative is defined by $\td_t \equiv \partial_t + \mathbf{v}_a \cdot \nabla$. The physical quantities associated with the individual species are the particle mass $m_a$ and charge $q_a$ while the fields are density $n_a$, pressure $p_a$, temperature $\bar{T}_a$ (in units of $k_{\text{B}}$) and velocity $\mathbf{v}_a$. Each species is assumed to satisfy the equation of state $p_a = n_a \bar{T}_a$. 

The particle dynamics is governed by the existence of a given \textit{external} background magnetic field $\mathbf{B} \equiv \mathbf{B}(\mathbf{x})$ created by some external steady current and satisfying $\nabla \cdot \mathbf{B} = 0$. The dynamical electromagnetic field satisfies the non-relativistic (electric limit) of the Maxwell equations (see App.~\ref{sec:nonrel} for details). In this limit there is no magnetic back-reaction on the particle dynamics. The magnetic field appearing in the Lorentz force of Eq.~\eqref{eq:fluidsystem} is therefore given entirely by the external background field $\mathbf{B}$. Although the magnetic part of the dynamical electromagnetic field exists in the non-relativistic limit (see Eq.~\eqref{eq:NRelectriclimit}), the field equation is decoupled from the rest of the dynamical system. The only coupling between the particle dynamics Eq.~\eqref{eq:fluidsystem} and the dynamical electromagnetic field is therefore through the Poisson equation for the electric potential,
\begin{align} \label{eq:emsystem}
\nabla^2 \phi = - \frac{\rho}{\varepsilon_0}  ~, 
\end{align}
where $\varepsilon_0$ is the electric (vacuum) permittivity. The electric potential appearing in Eq.~\eqref{eq:emsystem} is the leading order in the non-relativistic expansion. With the addition of initial and boundary conditions Eq.~\eqref{eq:fluidsystem} and Eq.~\eqref{eq:emsystem} provides a coupled set of equations for the dynamical fields $n_a, \mathbf{v}_a$ and $\phi$ which are driven by the external magnetic field $\mathbf{B}$. 

For clarity of the presentation, we will in this work present the above system in the case of two species of charged particles, electrons and ions with atomic number $Z=1$, labeled by $a=\{\text{e}, \text{i}\}$, respectively. The system of equations we consider is however straight-forwardly generalizable to the case of multiple species. The charge density and current density are given by
\begin{equation} \label{eq:chargecurrent}
 \rho = q_{\text{i}} n_{\text{i}} + q_{\text{e}} n_{\text{e}} ~, \quad
 \mathbf{J} = n_{\text{i}} q_{\text{i}} \mathbf{v}_{\text{i}} + n_{\text{e}} q_{\text{e}} \mathbf{v}_{\text{e}}~.
\end{equation}
Charge-current conservation is therefore directly satisfied by the sum of the continuity equations. We will consider the system \eqref{eq:fluidsystem}-\eqref{eq:chargecurrent} in a perturbative setup where the fields are assumed to approximate the solution in the form of perturbative expansions. It is therefore convenient to first put the system in dimensionless form. Introducing the characteristic length $\ell$, frequency $\omega$, and velocity $\bar{v} = \omega\ell$ together with the scale of the magnetic field $\bar{B}$, density $\bar{n}$, and temperatures $\bar{T}_a$ with $\bar{T}_{\text{e}} = m_{\text{i}} \bar{v}^2$, the system in dimensionless form is
\begin{align} \label{eq:normalized} 
\begin{split}
	\varepsilon \lambda \nabla^2 \phi &= n_{\text{e}} - n_{\text{i}}  ~, \quad
	\partial_t n_a + \nabla \cdot ( n_a \mathbf{v}_a ) = 0 ~, \\
	\varepsilon \mu_a \td_t \mathbf{v}_a 	
	&= - \varepsilon \tau_a\nabla \log( n_a ) + z_a ( - \nabla \phi + \mathbf{v}_a \times \mathbf{B} ) ~,
\end{split}
\end{align}
where all quantities are now dimensionless. The parameter $z_a$ specifies the sign and multiple of the elementary charge for the species and the dimensionless parameters are $\mu_a = \frac{m_a}{m_{\text{i}}}$, $\tau_a = \frac{\bar{T}_{a}}{\bar{T}_{\text{e}}}$, $\varepsilon = \frac{\omega}{\omega_{\text{c,i}}}$ and $\lambda = \frac{\omega_{\text{c,i}}^2}{\omega_{\text{p,i}}^2} = \frac{\varepsilon_0 \bar{B}^2}{m_{\text{i}} \bar{n}}$ with the ion cyclotron frequency $\omega_{c,\text{i}} = \frac{e \bar{B}}{m_i}$ and ion plasma frequency $\omega_{p,\text{i}}^2 = \frac{\bar{n} e^2}{\varepsilon_0 m_{\text{i}} }$. The physical parameters are the mass ratio $\mu_a$ and the temperature ratio $\tau_a$ together with the initial and boundary conditions, while the remaining parameters $\varepsilon$ and $\lambda$ simply set the units. In particular, if $\omega$ is chosen to be $\omega_{c,\text{i}}$, the characteristic length is $\ell = \rho_s \equiv \frac{\sqrt{m_{\text{i}} \bar{T}_{\text{e}}}}{e\bar{B}}$ such that the characteristic velocity $\bar{v}$ is the cold ion acoustic speed and $\lambda = \frac{\lambda_{\text{D}}^2}{\rho_s^{2}}$ with the Debye length $\lambda_{\text{D}}^2 = \frac{\varepsilon_0 \bar{T}_{\text{e}}}{\bar{n} e^2}$. The dynamical system \eqref{eq:normalized} describes a non-relativistic fluid of charged particles. It therefore has a thermodynamical description that naturally gives rise to a number of conserved quantities. In App.~\ref{sec:charges}, we provide the associated conserved quantities and show how the total energy can be obtained from the fluid energy conservation equation.

\subsection{Perturbative setup} \label{sec:scheme2}

We will in the following assume non-zero positive fluid densities and effectively consider two spatial dimensions by assuming that the direction of the external magnetic field $\mathbf{b} = \frac{\mathbf{B}}{B}$ is a preferred \emph{constant} direction. Here, we have defined $B \equiv \|\mathbf{B}\|_2$. We will for simplicity neglect the dynamics in the direction of the external magnetic field. The parallel component $v_{a,\parallel} = \mathbf{b} \cdot \mathbf{v}_a$ of the velocity is therefore assumed to vanish.

In the case where the dynamics of the densities $n_a$ is much slower than the dynamics of the velocity fields $\mathbf{v}_a$, the system \eqref{eq:normalized} exhibits two time scales. It is therefore natural to split the system into two subsystems, one for the fast dynamics and another for the slow dynamics. This is possible in the framework of singular perturbation methods. For many practical purposes, we do not require to resolve the fast particle dynamics, i.e. the physics on the scale of the ion gyro-frequency. We thus consider a long-wavelength expansion where all derivatives come with a factor of $\varepsilon$. In addition, we are mainly interested in the systems where the densities are sufficiently large for the system to be in a near neutral state. Note that in the limit $\lambda = 0$, the system of equations \eqref{eq:normalized} are invariant under scaling transformations of the density. We thus consider the two-fluid system \eqref{eq:normalized} in a perturbation in $\varepsilon$ while keeping $\lambda$, $\mu_a$ and $\tau_a$ finite.\footnote{For example, for $\bar{B}=1 \text{T}$ and $\bar{n} = 10^{19} \text{m}^{-3}$, one has $\lambda \approx \mu_{\text{e}}$.} The perturbative equations are equivalently obtained by choosing $\omega = \omega_{c,\text{i}}$ and scaling the independent variables by $t \rightarrow\varepsilon^{-1}t$ and $\mathbf{x} \rightarrow \varepsilon^{-1} \mathbf{x}$, together with the scaling of the electric potential $\phi \rightarrow\varepsilon^{-1} \phi$ such that the small scale structures of the electrical potential are suppressed. Other choices of perturbation setups are discussed in App.~\ref{sec:alternativeperb}.

The particular placement of the expansion parameter $\varepsilon$ in Eq.~\eqref{eq:normalized} puts both Poisson's equation and the momentum equation into the form of a singular perturbation problem, where the \zeroth-order equation has a lower differential order than the unperturbed equation. This means that the degenerate system can not satisfy all initial and boundary conditions in general. In particular, the velocity fields $\mathbf{v}_a$ are no longer dynamical and their initial conditions are specified by the initial conditions for the densities $n_a$ and electric potential $\phi$. We nevertheless expect that a solution to the full system \eqref{eq:normalized} will approach the solution of the perturbative system much in the same way ordinary dynamical systems behave \cite{Tihonov} (at least for some initial conditions and for some time interval). Under this assumption, we can therefore restrict to considering the singular perturbation problem and consider the \textit{dynamical} fields in a formal power series,
\begin{align} \label{eq:series}
n_a = \sum_{k=0}^{\infty} \varepsilon^{k} n_a^{(k)}, \quad \phi = \sum_{k=0}^{\infty} \varepsilon^{k} \phi^{(k)}, \quad \mathbf{v}_a = \sum_{k=0}^{\infty} \varepsilon^{k} \mathbf{v}_a^{(k)} ~.
\end{align}
Since we can not hope to prove convergence, we will simply assume that the expansion behaves as a derivative expansion and consider the first two terms. The best we could do would be to estimate the radius of convergence by extrapolation using a finite number of terms. In the end of the computation, we of course set $\varepsilon$ to unity and keep in mind that the expressions only hold as a long-wavelength, low frequency expansion i.e. for sufficiently slowly varying configurations. We thus consider the system \eqref{eq:normalized} order-by-order in $\varepsilon$.

Due to the singular perturbative nature of the momentum equation, we can solve for the components of the velocities perpendicular to the direction of the external magnetic field. The first couple of orders of the perpendicular components of the fluid velocities are\footnote{In App.~\ref{sec:alternativeperb}, we discuss perturbative setups where the diamagnetic-drift velocity appear at \zeroth order.}
\begin{align} \label{eq:vel012}
	\mathbf{v}^{(0)}_{a,\perp} = \mathbf{v}^{(0)}_{\text{E}} ~,\quad
	\mathbf{v}^{(1)}_{a,\perp} = \mathbf{v}^{(1)}_{\text{E}} + \mathbf{v}^{(0)}_{\text{D}a} + \mathbf{v}^{(0)}_{\text{P}a} ~,	 \quad \mathbf{v}^{(2)}_{a,\perp} = \mathbf{v}^{(2)}_{\text{E}} + \mathbf{v}_{\text{D}a}^{(1)} + \mathbf{v}_{\text{P}a}^{(1)} ~.
\end{align}
with the perpendicular drift velocities defined by
\begin{align}
	\mathbf{v}^{(k)}_{\text{E}} &\equiv \frac{\mathbf{b} \times \nabla \phi^{(k)}}{B} ~, \quad
	\mathbf{v}_{\text{D}a}^{(0)} \equiv \tau_{a} \frac{\mathbf{b} \times \nabla \log n^{(0)}}{z_a B}  ~, \quad \label{eq:drifts}
	\mathbf{v}_{\text{P}a}^{(0)} \equiv \mu_{a} \frac{\mathbf{b} \times \td^{(0)}_t \mathbf{v}^{(0)}_{\text{E}}}{z_a B} ~, \\
	\mathbf{v}_{\text{D}a}^{(1)} &\equiv \frac{\tau_a}{z_a n^{(0)}B} \mathbf{b} \times \left( \nabla n^{(1)}_a - n^{(1)}_a \nabla \log n^{(0)} \right) ~, \quad
	\mathbf{v}_{\text{P}a}^{(1)} \equiv \frac{\mu_a}{z_a B} \mathbf{b} \times \left( \td_t^{(0)} \mathbf{v}^{(1)}_{a,\perp} + \mathbf{v}^{(1)}_{a,\perp} \cdot \nabla \mathbf{v}^{(0)}_{\text{E}} \right)	~. \nonumber
\end{align}
One might expect that the electron-polarization drift is negligible compared to the ion-polarization drift as it is suppressed with a factor of $\mu_{\text{e}}$. However, for completeness we will keep it throughout. We will drop the perpendicular subscript in the following. 

At \zeroth order, the Poisson equation is simply $\rho^{(0)} = 0$ and the system is therefore neutral. In fact, this implies that any charge separation in the system that might occur, e.g. polarization or diamagnetic velocity drifts must enter at a higher order. Consequential, the velocities must be equal $\mathbf{v}^{(0)}_{\text{i}} = \mathbf{v}^{(0)}_{\text{e}}$ and therefore $\mathbf{J}^{(0)} = 0$. We can now define a single fluid density $n^{(0)} = n^{(0)}_{\text{i}} = n^{(0)}_{\text{e}}$ (at this order) and since the parallel component of the velocity fields is neglected write the continuity equation,
\begin{align} \label{eq:continuity0}
\partial_t n^{(0)} + \nabla \cdot \mathbf{\Gamma}^{(0)} = 0 ~.
\end{align}
where we have introduced the density flux $\mathbf{\Gamma}^{(0)}= n^{(0)} \mathbf{v}^{(0)}_{\text{E}}$. The system of \zeroth order equations is closed by the \first-order Poisson equation $\lambda \nabla^2 \phi^{(0)} = - \rho^{(1)}$ and the corresponding charge-current density conservation equation,
\begin{align} \label{eq:chargecurrentdensity}
\partial_t \rho^{(1)} + \nabla \cdot \mathbf{J}^{(1)} = 0 ~.
\end{align}
The \first-order correction to the current is $\mathbf{J}^{(1)} = n^{(0)} (\mathbf{v}_{\text{D}}^{(0)} + \mathbf{v}_{\text{P}}^{(0)}) + \rho^{(1)}\mathbf{v}_{\text{E}}^{(0)}$ with the combined diamagnetic and polarization-drift velocities $\mathbf{v}_{\text{D}}^{(0)} = \mathbf{v}_{\text{Di}}^{(0)} - \mathbf{v}_{\text{De}}^{(0)}$ and $\mathbf{v}_{\text{P}}^{(0)} = \mathbf{v}_{\text{Pi}}^{(0)} - \mathbf{v}_{\text{Pe}}^{(0)}$, respectively. We note in passing that in the case of multiple species, the ion density $n_{\text{i}}^{(0)}$ can simply be replaced by the total density of (positive) ions such that factors of $\mathbf{v}^{(1)}_{\text{E}}$ does not appear in the current.

Due to the singular drift expansion of the perpendicular velocities \eqref{eq:vel012}, time derivatives are introduced in the current $\mathbf{J}^{(1)}$ itself by the polarization drift. In fact, this is the reason that the seemingly strict neutral approximation $\lambda = 0$, in which Poisson's equation $n^{(k)}_{\text{i}} = n^{(k)}_{\text{e}}$, can make sense even though it is equivalent to imposing $\nabla \cdot \mathbf{J}^{(k)} = 0$. This is exactly what is meant by the quasi-neutral approximation. To evolve the charge density $\rho^{(1)}$ both the electric potential $\phi^{(0)}$ and its time derivative $\partial_t \phi^{(0)}$ have to be evaluated at equal times. Using the Poisson equation and the charge current conservation equations \eqref{eq:chargecurrentdensity} one obtains the elliptic equation
\begin{equation} \label{eq:dphit}
\mathcal{D} \left[ \partial_t \phi^{(0)} \right] + \nabla \cdot \tilde{\mathbf{J}}^{(1)} = 0~.
\end{equation}
where we have introduced the modified current $\tilde{\mathbf{J}}^{(1)} = n^{(0)} (\mathbf{v}^{(0)}_{\text{D}} + \mathbf{v}^{(0)}_{\text{S}}) + \rho^{(1)}\mathbf{v}_{\text{E}}^{(0)}$ with $\mathbf{v}^{(0)}_{\text{S}} = - (\mu_{\text{i}} + \mu_{\text{e}}) \frac{ 1 }{B} \left( \mathbf{v}_{\text{E}}^{(0)} \cdot \nabla \right) \frac{ \nabla_{\perp} \phi^{(0)} }{B}$ and the differential operator $\mathcal{D} \equiv - \nabla \cdot \left( (\mu_{\text{i}} + \mu_{\text{e}}) \frac{n^{(0)} }{B^2}  + \lambda \right) \nabla_{\perp}$ using that $\nabla \phi = \nabla_{\perp} \phi$ with $\nabla_{\perp} = \mathbf{b} \times (\nabla \times \mathbf{b})$. The boundary value problem given by Eq.~\eqref{eq:dphit} requires an independent set of boundary conditions for the time-derivative of the $\phi^{(0)}$ to be specified. The two equations \eqref{eq:continuity0} and \eqref{eq:dphit} provide a consistent system for the fields $n^{(0)}$ $\phi^{(0)}$, $\partial_t \phi^{(0)}$ through the corresponding source $\tilde{\mathbf{J}}^{(1)}$. The dynamics at the lowest order is thus driven by the charge separating terms $\mathbf{v}^{(0)}_{\text{D}a}$ and $\mathbf{v}^{(0)}_{\text{P}a}$ even though they are in principle \first-order quantities.

At this point it is worth commenting on a couple of details of how the perturbative equations comes out. First, we note that it is sufficient to solve the elliptic equation \eqref{eq:dphit} and simply treat $\phi^{(0)}$ as a dynamical field (in place of solving the Poisson equation). This comes at the cost of exchanging the boundary conditions for $\phi^{(0)}$ with initial conditions . Proceeding to the next order, where the \second-order time derivative of the electric potential is required, we note that even $\partial_t \phi^{(0)}$ can be treated as a dynamical field. In this case the boundary conditions for $\partial_t \phi^{(0)}$ are traded with an additional set of (dependent) initial conditions. In this way, each order only introduces one additional elliptic equation. Secondly, the set of equations at a given order is only closed by using information about the velocity field from the succeeding order. This is generic for all orders of this particular perturbative setup and makes the diamagnetic and polarization-drift velocities as well as the charge and current densities of the next order available in terms of the lower order quantities. In this sense, Eq.~\eqref{eq:chargecurrentdensity} is redundant for obtaining the \zeroth-order fields. Finally, there exists an alternative (but equivalent) set of equations that takes the place of  Eq.~\eqref{eq:dphit} at the cost of introducing an auxiliary field (see e.g. \cite{doi:10.1063/1.4985318} for the \zeroth order).
The presented set of equations \eqref{eq:continuity0}--\eqref{eq:dphit} have the advantages that they have a straight-forward generalization to systems with multiple participating species. 

The \first-order corrections to the densities can be computed from the continuity equations
\begin{align} \label{eq:continuity1}
\partial_t n_a^{(1)} + \nabla \cdot \mathbf{\Gamma}^{(1)}_a = 0 ~,
\end{align}
where the correction to the flux density is $\mathbf{\Gamma}^{(1)}_a = n^{(0)} \mathbf{v}_a^{(1)} + n_a^{(1)} \mathbf{v}_{\text{E}}^{(0)}$. Note that $\mathbf{J}^{(1)} = \mathbf{\Gamma}^{(1)}_{\text{i}} - \mathbf{\Gamma}^{(1)}_{\text{e}}$. The system of \first-order equations is closed by the Poisson equation $\lambda \nabla^2 \phi^{(1)} = - \rho^{(2)}$ and the \second-order charge-current density conservation
\begin{align} \label{eq:chargecurrentdensity1}
\partial_t \rho^{(2)} + \nabla \cdot \mathbf{J}^{(2)} = 0 ~.
\end{align}
The \second-order correction to the current is $\mathbf{J}^{(2)} = n^{(0)} (\mathbf{v}_{\text{D}}^{(1)} + \mathbf{v}_{\text{P}}^{(1)}) + n^{(1)}_{\text{i}} \mathbf{v}^{(1)}_{\text{i}} - n^{(1)}_{\text{e}} \mathbf{v}^{(1)}_{\text{e}} + \rho^{(2)} \mathbf{v}_{\text{E}}^{(0)}$ with the combined diamagnetic and polarization-drift velocities $\mathbf{v}_{\text{D}}^{(1)} = \mathbf{v}_{\text{Di}}^{(1)} - \mathbf{v}_{\text{De}}^{(1)}$ and $\mathbf{v}_{\text{P}}^{(1)} = \mathbf{v}_{\text{Pi}}^{(1)} - \mathbf{v}_{\text{Pe}}^{(1)}$, respectively. As a consequence of the singular scheme, the \second-order correction to the current $\mathbf{J}^{(2)}$ now involves both the time derivative of the correction to the electric potential $\partial_t \phi^{(1)}$ and the second-order derivative $\partial^2_t \phi^{(0)}$. An equation for the latter is obtained directly from Eq.~\eqref{eq:dphit} while an equation for the first is obtained in a similar way to Eq.~\eqref{eq:dphit}, i.e. we take the combination of the \first-order Poisson and charge-current equation \eqref{eq:chargecurrentdensity1},
\begin{equation} \label{eq:dphi1t}
\mathcal{D} \left[ \partial_t \phi^{(1)} \right] + \nabla \cdot \tilde{\mathbf{J}}^{(2)} = 0 ~,\quad
\mathcal{D} \left[ \partial^2_t \phi^{(0)} \right] + \nabla \cdot \left(\partial_t \tilde{\mathbf{J}}^{(1)}\right) = 0
\end{equation}
with the modified current $\tilde{\mathbf{J}}^{(2)} = n^{(0)}(\mathbf{v}_{\text{D}}^{(1)} + \mathbf{v}_{\text{S}}^{(1)}) + n^{(1)}_{\text{i}} \mathbf{v}^{(1)}_{\text{i}} - n^{(1)}_{\text{e}} \mathbf{v}^{(1)}_{\text{e}} + \rho^{(2)} \mathbf{v}_{\text{E}}^{(0)}$ with 
$\mathbf{v}_{\text{S}}^{(1)} = \frac{\mathbf{b}}{B} \times (\partial_t(\mu_{\text{i}}(\mathbf{v}_{\text{Di}}^{(0)} + \mathbf{v}_{\text{Pi}}^{(0)}) + \mu_{\text{e}}(\mathbf{v}_{\text{De}}^{(0)} + \mathbf{v}_{\text{Pe}}^{(0)})) + \mathbf{v}_{\text{E}}^{(0)} \cdot \nabla (\mu_{\text{i}}\mathbf{v}^{(1)}_{\text{i}}+\mu_{\text{e}}\mathbf{v}^{(1)}_{\text{e}}) + (\mu_{\text{i}}\mathbf{v}^{(1)}_{\text{i}}+\mu_{\text{e}}\mathbf{v}^{(1)}_{\text{e}})\cdot \nabla \mathbf{v}_{\text{E}}^{(0)})$.
The second equation is obtained from Eq.~\eqref{eq:dphit}, it is therefore an equation in terms of \zeroth-order fields $\partial_t \tilde{\mathbf{J}}^{(1)} = n^{(0)}\partial_t(\mathbf{v}_{\text{D}}^{(0)} + \mathbf{v}_{\text{S}}^{(0)}) + \rho^{(1)} \partial_t\mathbf{v}_{\text{E}}^{(0)}  - \nabla\cdot\mathbf{J}^{(1)}\mathbf{v}_{\text{E}}^{(0)} - \nabla\cdot\mathbf{\Gamma}^{(0)} (\mathbf{v}_{\text{D}}^{(0)} + \mathbf{v}_{\text{P}}^{(0)})$. With Eq.\eqref{eq:dphi1t}, it is possible to treat $\phi^{(1)}$ and $\partial_t \phi^{(0)}$ as dynamical fields at the cost of exchanging their boundary conditions with consistent initial conditions. The complete perturbative system of equations to \first order is thus given by Eq.~\eqref{eq:continuity0}-\eqref{eq:chargecurrentdensity}, Eq.~\eqref{eq:continuity1} and Eq.~\eqref{eq:chargecurrentdensity1} together with the two elliptic problems given by Eq.~\eqref{eq:dphi1t} and the dynamic evolution of the fields $\partial_t \phi^{(0)}$, $\phi^{(0)}$, and $\phi^{(1)}$. Notice that similarly to the preceding order, the \second-order corrections to the charge and current densities are available through Eq.~\eqref{eq:chargecurrentdensity1} in terms of only \first-order quantities.

\subsubsection*{Conserved quantities}

A particular placement of the perturbation parameter $\varepsilon$ can potentially break the conservation of a conserved charge of the full system,\footnote{For example by splitting the convection term in a conservation equation.} such that it is conserved only if all orders are included. We therefore check if the perturbation equations retains the conservation of the constants of motion at each order. The conserved charges of the full system are given in App.~\ref{sec:charges}. 

If contributions from the boundary vanish (i.e. vanishing fluxes at the boundary), the conservative form of Eq.~\eqref{eq:continuity0}-\eqref{eq:chargecurrentdensity} and Eq.~\eqref{eq:continuity1}-\eqref{eq:chargecurrentdensity1} ensures that mass and charge conservation is preserved at each order. The integrated fields $n^{(0)}$, $n_a^{(1)}$, $\rho^{(1)}$, $\rho^{(2)}$ are therefore constants of motion. In this case, the total energy $E$ is therefore also preserved at each order. Assuming that the fields vanish at the boundary, the first couple of terms in the expansion of the total energy density are given by
\begin{align}
E^{(0)} =&\; \frac{\lambda}{2} \left(\nabla \phi^{(0)}\right)^2 + \sum_{a} n^{(0)} \left( \tau_{a} \log n^{(0)} + \frac{\mu_{a}}{2} \frac{\left(\nabla \phi^{(0)}\right)^2}{B^2} \right) ~, \label{eq:E0perb} \\
\begin{split}
E^{(1)} =&\; \lambda \nabla \phi^{(0)} \cdot \nabla \phi^{(1)} + \sum_{a} \tau_{a} n^{(1)}_a \left(\log n^{(0)} + 1 \right) + \\
& \frac{\mu_{a}}{B^2}\nabla \phi^{(0)} \cdot \left( \frac{1}{2} n^{(1)}_a \nabla \phi^{(0)} + n^{(0)} \left( \nabla \phi^{(1)} + \frac{\tau_a}{z_a}\nabla \log n^{(0)} + \frac{\mu_a}{z_a} \td^{(0)}_t \mathbf{v}^{(0)}_{\text{E}}  \right)\right) \label{eq:E1perb} ~.
\end{split}
\end{align}
These can be obtained either directly by inserting the expansion into Eq.~\eqref{eq:totalenergy} or by using the equations of motion. The lowest contribution $E^{(0)}$ given by Eq.~\eqref{eq:E0perb} is obtained by multiplying Eq.~\eqref{eq:dphit} by $\phi^{(0)}$, integrating over space and using integration by parts. This will leave two residual terms, one can be exchanged by the considering the volume integral over $\tau_a \partial_t(n^{(0)} \log n^{(0)})$ and using the continuity equation \eqref{eq:continuity0}, the other by using the Poisson equation. In a similar fashion, the \first-order correction $E^{(1)}$ given by Eq.~\eqref{eq:E1perb} can be found by considering Eq.~\eqref{eq:dphi1t} and using the remaining equations of motion.
 
\subsubsection*{Discussion}

Some general comments on the presented perturbative equations can now be made.

In many cases the gradient of the density is substantial and it would seemingly be preferable to treat the diamagnetic-drift velocity at the same order as the $\text{E}{\times}\text{B}$-drift velocity. In this regard, the velocity expansion given by Eq.~\eqref{eq:vel012} is seemingly an unconventional drift-ordering (see e.g. \cite{doi:10.1063/1.4943199}). However, this requires a perturbative setup where charge separation can be accommodated at \zeroth order. This means that the separation of charge is assumed to be sufficiently large as to be treated as a \zeroth-order quantity. But in such cases, a quasi-neutral limit $\lambda = 0$ can not make sense. It is therefore particularly interesting to consider the different limits of the parameter $\lambda$ which sets the scale of the density, i.e. $\lambda \propto \frac{\bar{B}^2}{\bar{n}}$. The strict quasi-neutrality limit $\lambda \rightarrow 0$ is reached for sufficiently large densities. In this limit, the system of equations \eqref{eq:normalized} becomes invariant under scale transformations of the density. The quasi-neutral system therefore predicts equivalent dynamics for all magnitudes of densities. On the other hand, for small densities $\lambda\sim\mathcal{O}(1)$, the small scale structure of the electric potential becomes increasingly important and charge separation can consistently be included at the lowest order. This always comes at the cost of removing the possibility of a quasi-neutral limit. In App.~\ref{sec:alternativeperb}, we discuss two such dilute perturbative setups. Even though the charge separating terms enters at first order in the perturbative setup presented in Sec.~\ref{sec:scheme2}, the diamagnetic and polarization-drift corrections for the next order are still available at any given order. It is therefore possible to obtain a better approximation for the velocity field by
\begin{equation} \label{eq:estvel}
	\mathbf{v}_a \approx \sum_k \tilde{\mathbf{v}}^{(k)}_a + \mathcal{O}(\mathbf{v}^{(k+1)}_{\text{E}}) \quad \text{with} \quad \tilde{\mathbf{v}}^{(k)}_a = \mathbf{v}^{(k)}_{\text{E}} + \mathbf{v}^{(k)}_{\text{D}a} + \mathbf{v}^{(k)}_{\text{P}a} ~.
\end{equation}

Naturally, the particle species appear in the system \eqref{eq:normalized} on equal footing; their dynamics distinguished by mass, charge and in the isothermal system by their constant temperature. As the sign of the carried charge and the mass ratio $\mu_{\text{e}}$ are fixed, the asymmetry between the dynamics of the species is solely controlled \emph{externally} by the ion-temperature ratio $\tau_{\text{i}}$. However, at \zeroth order given by Eq.~\eqref{eq:continuity0}-\eqref{eq:dphit} the dynamics is fully determined by the combined parameters $\mu_{\text{i}} + \mu_{\text{e}}$ and $\tau_{\text{i}} + \tau_{\text{e}}$ multiplying the combined-drift velocities $\mathbf{v}^{(0)}_{\text{P}}$ and $\mathbf{v}^{(0)}_{\text{D}}$, respectively. At this order, the system therefore does not distinguish between the species. The asymmetry between the species appears at \first order through the mass ratio $\mu_{\text{e}}$ as the \first-order combined-drift polarization velocity $\mathbf{v}^{(1)}_{\text{P}}$ now introduces cross-terms with the polarization-drift velocity, $\mu_{\text{i}}^2 - \mu_{\text{e}}^2 \approx \mu_{\text{i}}^2$ and the diamagnetic-drift velocity $\mu_{\text{i}}\tau_{\text{i}} - \mu_{\text{e}}\tau_{\text{e}} \approx \mu_{\text{i}}\tau_{\text{i}}$. One therefore expects that non-zero ion temperature plays a significant role for the \first-order corrections.

\section{Simulation} \label{sec:simulation}

We perform a quantitative analysis of the perturbative equations given by Eqs.~\eqref{eq:continuity0}-\eqref{eq:dphi1t} using numerical methods. In particular, we are interested in what parameter regimes the perturbation series converges to a degree where we can trust that it represents the dynamics of the full system given by Eq.~\eqref{eq:normalized}. This is possible, since we have direct access to the first-order corrections of the dynamical fields in a numerical simulation. We can thus monitor when the corrections become too important to say that the lowest order represent a converged series on its own. In order to control the different dynamical features of the two-fluid we consider the simulated dynamics of a seeded blob using a numerical implementation of the perturbation equations. Here, we are mainly interested in the validity of the perturbative equations and are not particular concerned with the blob dynamics itself. We will therefore restrict to considering the fields themselves and not the center of mass quantities. A great deal of work can be found in the literature on the dynamics of seeded blobs (see e.g. \cite{HeldBlob} and references within). The numerical implementation and convergence of the numerical methods have been verified using the method of manufactured solutions. For actual simulations, we have used the conserved charges, i.e. mass, charge and energy of the system to monitor if the numerical solution stays within physical constraints.

Let us now define the operator $C^{(n)}_{p}(f) \equiv \frac{\|f^{(n)}\|_p}{\|f^{(n-1)}\|_p}$ with $n,p \in \mathbb{N}_{+}$ that gives the $L^p$-norm ratio of the $n$'th order field coefficient to the $(n-1)$'th order field coefficient. We will use this operator to estimate the importance of the higher-order corrections. Furthermore, we introduce the vorticity $\boldsymbol{\omega}_a \equiv \nabla \times \mathbf{v}_a$ with the parallel component $\omega_a \equiv \mathbf{b} \cdot \boldsymbol{\omega}_a$.

\subsection{Initial conditions}

We choose a Cartesian coordinate system $\{t,x,y,z\}$ with the direction of the magnetic field set to $\mathbf{b} = \mathbf{e}_z$. The magnitude of the external magnetic field is set to $B^{-1}(x) = 1 + \kappa(x-x_0)$, where the curvature is parametrized by $\kappa$ satisfying the positivity condition $\kappa x_0 < 1$. The dynamical fields are put on a discrete two-dimensional grid spanned by the $x$ and $y$ coordinates. We run simulations with a seeded blob on a domain $\mathcal{D}$ with periodic boundary conditions in the $y$-direction and Dirichlet conditions in the $x$-direction. The blob is initialized in the \zeroth-order density $n^{(0)}$ by a two-dimensional Gaussian function
\begin{align} \label{eq:ICn}
	n^{(0)} = 1 + A \exp\left(-\frac{(\mathbf{x} - \mathbf{x}_0)^2}{2\sigma^2}\right) ~,\quad n_a^{(k)} = 0 ~,
\end{align}
with amplitude $A$, equal width $\sigma$ in both directions and its initial center of mass at $\mathbf{x}=\mathbf{x}_0 \in \mathcal{D}$. We note that the system is by construction neutral at \zeroth order and the initial conditions are compatible with the perturbative assumption $C_p^{(1)}(n_a) \ll 1$. The boundary conditions for the densities are	$n^{(0)} |_{\partial\mathcal{D}}  \approx 1$ and $n_a^{(k)} |_{\partial\mathcal{D}} = 0$. If the total area of the simulation domain is denoted by $V$, the total density in the system is $\int dV n \approx \bar{n} (V + 2\pi A\sigma^2)$. The electric potential $\phi$ (to all orders) must be zero initially because of the initial conditions for the densities. 

The perpendicular components of the velocities are given completely in terms of the density and electric potential due to the (singular) drift expansion \eqref{eq:vel012}. Their initial and boundary conditions are therefore also determined by the initial and boundary conditions of the dynamical fields. The choice of initial conditions must therefore be chosen with care. While the above specified conditions innocuously sets the blob at rest (to lowest order) initially, the \zeroth-order velocity can not possibly resemble the true velocity field at initial times $t_0$ (as the correction completely dominates the \zeroth order, i.e. $C_p^{(1)}(\mathbf{v}_a) \gg 1$). However, since the system is invariant under time-reversal, a simulation can be continued from any point in time. Thus, if the system fairly quickly orders itself, a \zeroth-order simulation initialized from rest can be thought of starting from a non-zero initial velocity at a later time $t_1$ where the \first-order corrections are sufficiently suppressed and the initial period can be discarded. The question then becomes whether the \quoting{evolved} initial condition is of interest. Thus if we consider a simulation on some finite interval $t_0 \leqslant t \leqslant T$, we expect rapidly changing velocity fields in a neighborhood of $t_0$, i.e. $t_0 \leqslant t \leqslant t_1$ and a transition to slowly changing fields (compared to the density and electric potential) in the remaining interval $t_1 \leqslant t \leqslant T$.

\subsection{Results}

The dynamics of a given solution is determined by the external parameter $\kappa$, the physical parameter $\tau_{\text{i}}$ representing the ratio between ion and electron temperature, the shape of the blob given by the parameters $A$ and $\sigma$ and finally the scale $\lambda\propto \frac{\bar{B}^2}{\bar{n}}$. The parameter $\lambda$ sets the scale of the density. Recall, that when the system is not strictly quasi-neutral ($\lambda \neq 0$), the system equations are not invariant under scaling transformations of the densities. We can therefore study the dynamics at different scales of the density. We consider the following regimes of parameter space (i) cold ions $\tau_{\text{i}}=0$ in the quasi-neutral limit $\lambda=0$, (ii) hot ions $\tau_{\text{i}}>0$ in the quasi-neutral limit $\lambda=0$, and (iii) the system at finite densities $\lambda>0$ (non-zero charge density). The first case varies the initial conditions and shows the significance of the spatial scales, while the latter two shows the significance of temporal scales by varying a physical parameter. In all cases we set the curvature parameter to $\kappa = 0.00015$.

\subsubsection*{Cold ions in the quasi-neutral limit}

We start by considering the dynamical case consisting of cold ions $\tau_{\text{i}}=0$ in the quasi-neutral limit $\lambda=0$ (infinite density limit). In this regime, the system of equations is density scale invariant and $n^{(k)}_{\text{e}} = n^{(k)}_{\text{i}}$. The dynamical fields for different parameter sets of $\{A, \sigma\}$ can therefore to some extent be related by scaling transformations even though the system is non-linear. The transformations can be estimated by using the initial conditions \eqref{eq:ICn} and the equations of motion. They should therefore at least hold at initial times. We omit the details on how these can be obtained and simply list the estimated scaling transformations for the operator $C$ applied on the fields in Tab.~\ref{tab:coldion}.
\input{coldiontable}

We now consider numerical simulations for a range of initial conditions $\{\frac{1}{4} \leqslant A \leqslant 4, 10 \leqslant \sigma \leqslant 70\}$ which sets the initial length scale of the density and thus determines the initial strength of the electron diamagnetic-drift velocity. In Fig.~\ref{fig:evolution0}, we show a typical evolution of the density field $n \approx n^{(0)} + n^{(1)}$ and the separate expansion coefficients.
\input{coldevolution}
The density is advected towards the right while increasing (decreasing) in the upper (lower) half plane, i.e. the blob moves slightly upwards. This is generally the case in the cold ion limit as the cross terms in $\tilde{\mathbf{J}}^{(2)}$ between the polarization and diamagnetic-drift velocities are dominated by the electrons. For sufficient non-zero ion temperature the roles are reversed and the blob moves downwards.
\input{sigmascaling}
In Fig.~\ref{fig:sigmascaling}, the velocity fields for different simulations are shown after applying the scaling transformations listed in Tab.~\ref{tab:coldion}. For a fixed amplitude $A$, the scaled fields agree within good approximation for all values of $\sigma$ on the entire time interval (see Fig.~\ref{fig:sigmascaling}(i,ii)). For varying amplitude $A$, the two velocity fields evolves on different time scales (see Fig.~\ref{fig:sigmascaling}(iii,iv)). So while the scaling relations provide almost perfect agreement for the electron velocity (as the diamagnetic-drift velocity dominates on the entire time interval), the ion velocity initially evolves on the same timescale for all values of the amplitude, but will after the initial acceleration grow at a faster rate for larger amplitudes. The density and electric potential are therefore also affected at later times albeit to a smaller degree (as the relative magnitude between the corrections for the two species are several orders). 

\input{cold0}
The operator $C$ applied on the components of the velocity fields for both the electron and ion species is shown in Fig.~\ref{fig:cold0} for the particular set $\{A=1, \sigma=50\}$. The \first-order corrections to the velocity fields decreases fairly rapidly over an initial period after which it becomes slowly varying. Since there are density gradients at all times, the \first-order correction to the electron velocity (which is dominated by the diamagnetic-drift velocity) is orders of magnitude larger than the correction to the ion velocity. For sufficiently small scales (i.e. small $\sigma$ or large amplitude $A$), the electron-velocity field is therefore not very well represented by the lowest order as is especially evident from the \zeroth-order vorticity field $\omega$ (see Fig.~\ref{fig:cold1}). However, this is not explicitly seen from the \first-order correction to the density (Fig.~\ref{fig:cold2}(i)) as the convection is to a large extent driven by the $\text{E}{\times}\text{B}$-drift velocity. Since the $\text{E}{\times}\text{B}$-convection in turn is driven by the sub-leading charge separating terms through the electric potential, it becomes necessary to consider the second-order corrections to the velocity in order to check the convergence. Fig.~\ref{fig:cold0}(ii) show the $C$ operator applied on the second-order corrections for the velocities. It shows that the \second-order corrections are almost of the same order as the \first-order correction to the density. Thus, including the \first-order correction in the electron velocity we see the same order of approximation.
\input{cold2}
Strictly speaking, it is necessary to check one order above the truncation-order to estimate the convergence of the given order of approximation. However, in the given perturbation setup we have access to higher order corrections to the velocity field at a given order. It therefore makes sense to include the terms in the velocity field at the given order (post simulation). This is exactly the corrected velocity field $\tilde{\mathbf{v}}_a$ given by Eq.~\eqref{eq:estvel}. In Fig.~\ref{fig:cold2}, the corrected velocity fields, density and electric potential are shown. The \first-order corrections to $\tilde{\mathbf{v}}_a$ is not surprisingly magnitudes smaller than the true \first-order velocity correction $\mathbf{v}^{(1)}_a$. In some sense, we can therefore use the corrected velocity to gain an estimate of how well the overall approximation is at a given order. However, it should be stressed that while the corrected velocity field gives a better approximation at a given order, it does not constitute the flux $\mathbf{\Gamma}$ at that same order.

In Fig.~\ref{fig:cold2}, all \first-order corrections are shown to grow with time. One way to interpret this is that when the density blob begins to spread out, small scales starts to develop as is particular prominent when the blob tail begins to spiral. Comparing the density field (Fig.~\ref{fig:evolution0}) with the magnitude of the \first-order correction to the vorticity (Fig.~\ref{fig:cold1}) by scaling time appropriately, we see that the onset of this local spinning is picked up by the vorticity field. In this case, the vorticity therefore provides a measure for the emergence of small scale structures. We also note that in the cold ion limit the relative vorticity can within good approximation be estimated in terms of lowest-order quantities by $\theta_a \equiv \frac{\|\omega^{(0)}_{\text{D}a}\|_2}{\|\omega^{(0)}_{\text{E}}\|_2}$. This quantity is sometimes also denoted the finite Larmor radius (FLR) strength parameter in the case of ions as it is used as a measure for the strength of the charge separation.
\input{cold1}
The numerical simulations and estimated scaling transformations (listed in Tab.~\ref{tab:coldion}) support that all higher-order corrections are suppressed at large spatial scales, i.e. with increasing $\sigma$ and decreasing amplitude $A$. As soon as the dynamics is considered on sufficiently small scales the corrections becomes too important to assert that the solution to the degenerate system has converged sufficiently to represent the true dynamics. The perturbation equations are therefore indeed valid as a long-wavelength expansion. In particular, we note that the \zeroth-order system given by Eqs.~\eqref{eq:continuity0}-\eqref{eq:dphit} can provide a converged solution in the cold ion limit (at least on some time interval) if the system is considered for sufficiently large scales and if the corrected velocity is used (post simulation) to approximate the velocity field.

\subsubsection*{Hot ions in the quasi-neutral limit}

We consider numerical simulations in the case of non-zero ion temperature $\tau_{\text{i}}$ in the quasi-neutral limit ($\lambda = 0$). In particular, we consider two parameter series, (i) one with varying ion temperature $\{A=1, \sigma=40, 0 \leqslant \tau_{\text{i}} \leqslant 4\}$, and (ii) one with varying width $\{A=1, \tau_{\text{i}}=1, 10 \leqslant \sigma \leqslant 70\}$. Similarly to the cold ion case, we list in Tab.~\ref{tab:hotion} scaling transformations based on the initial conditions \eqref{eq:ICn} and the equations of motion. The transformations relates the dynamical fields for different $\sigma$ and $\tau_{\text{i}}$ and holds for sufficiently large ion-temperature ratio $\tau_{\text{i}}$.
\input{hotiontable}

For non-zero ion temperature the asymmetry between the species becomes much more apparent in the dynamics as the cross terms between the polarization and diamagnetic-drift velocities are much larger due to the ion mass ratio $\mu_{\text{i}}$ being unity. We therefore observe a transition point in the dynamics with increasing ion temperature when the cross terms are in balance. At sufficiently large ion-temperature, this balance is broken and we leave the cold ion limit. In this case, the blob has a downwards movement. An example of this is shown in Fig.~\ref{fig:evolution1}. In this regime, the dynamics happens on a faster time scale (compared to the cold ion limit, see e.g. Fig.~\ref{fig:evolution0}). The faster movement can therefore result in the blob leaving behind a depletion in the density which can only be \quoting{filled out} by including higher-order corrections. For the \first-order system this explicitly shows up when the corrections to the density becomes as large as the \zeroth order and a negative density is obtained. In Fig.~\ref{fig:evolution1}, this is seen at the last time stamp.
\input{hotevolution}
The faster dynamics is a direct consequence of the larger charge separation in the system at non-zero ion temperature which generates a larger electric potential. At some point in time, the $\text{E}{\times}\text{B}$-drift velocity therefore grows at a much faster rate which is reflected directly in the density movement. Whether the \first-order solution gives such large corrections depends on the parameters as can be seen from Fig.~\ref{fig:nInf}. In particular, there exist a critical parameter domain for which the \first-order system inevitably will loose its predictive power as the \first-order correction becomes comparable with the \zeroth order. In this parameter domain higher-order corrections are necessary for the series expansion to give a meaningful approximation. We note that this statement refers to the uniform convergence of the solution. Looking at the overall density field, the corrections are very localized in space and the corrections can therefore be suppressed by increasing the blob width.
\input{tausigma}
In Fig.~\ref{fig:hot0}, the \first-order coefficient $C^{(1)}_2$ is shown for the density and corrected velocity fields for the particular parameters $\{\sigma = 40, \tau_{\text{i}}=\frac{1}{2}\}$. It is clear that the correction to the velocity field grows with the ion temperature (comparing with Fig.~\ref{fig:cold2}). Nevertheless, for low ion temperatures and blob width, the \zeroth-order approximation can provide a fairly converged solution. For moderate ion temperatures and blob width (see Fig.~\ref{fig:hot0}(ii)), the \first-order corrections becomes necessary as the \zeroth-order approximation can not capture the dynamics single-handed. Finally, for high ion temperatures, the corrections to the velocity field becomes comparable with the \zeroth-order approximation and higher-order corrections are required. Also, it is interesting to note that due to the large $\text{E}{\times}\text{B}$-drift velocity, the corrected velocity field $\tilde{\mathbf{v}}_a$ almost align for both species  when the ion temperature ratio is near unity even though the diamagnetic and polarization-drift velocities come with opposite contributions for the two species.
\input{hot0}
Naturally, the correction to both vorticity fields also increases with the ion temperature with the electrons still having much faster local spinning. In particular, $\theta_{\text{e}}$ decreases with increasing ion temperature and provides a poor estimate of the relative electron vorticity. On the other hand, the FLR parameter for the ions $\theta_{\text{i}}$ agrees to a much larger degree with the relative ion vorticity at moderate temperatures

The numerical simulations for various ion temperature together with the scaling transformations given in Tab.~\ref{tab:hotion} show that the perturbation equations to first order are valid for low frequencies and second or higher-order corrections are necessary to resolve the faster dynamics. In addition, the dependence on $\sigma$ reveals (as for the cold ion limit) that the perturbation equations are valid as a long-wavelength expansion.

\subsubsection*{Finite density limit}

We consider numerical simulations for finite densities by considering non-zero positive $\lambda$ for a fixed set of initial conditions $\{A=1, \sigma=20\}$. We consider two parameter series, (i) one for vanishing ion temperature $\tau_{\text{i}}=0$, and (ii) one for unity ion-electron temperature ratio $\tau_{\text{i}}=1$. For a non-zero $\lambda$, the Poisson equation enters the system of equations and the scale invariance of the system is broken. The parameter $\lambda$ is thus a measure for the significance of the Poisson equation. For finite $\lambda$, the perturbation equations still requires the system to be neutral at the lowest order (by construction), i.e. $\rho^{(0)}=0$. The charge density therefore evolves due to the split in \first-order density corrections. With increasing $\lambda$, the electric potential is determined progressively more by the charge density instead of the charge separating diamagnetic and polarization-drift velocities. As a consequence, the dynamics happens on a much slower time scale with increasing $\lambda$.
\input{lambdatau0}
In Fig.~\ref{fig:chargedensity}(i), the relative size between the \second-order correction and the \first-order correction to the charge density is shown in the cold ion limit $\tau_{\text{i}}=0$ for different values of $\lambda$. The significance of the \second-order correction increases as the density becomes more dilute. Already for $\lambda = 1$ corresponding to $\bar{n} \approx 10^{15} m^{-3}$ at $\bar{B} = 1T$, we observe quite significant corrections. For sufficiently dilute systems, the \first-order corrected charge density (i.e. $\rho = \rho^{(0)} + \rho^{(1)}$) therefore fails to converge. For non-zero ion temperature, there is a balance between the contribution from the charge separating velocities and the charge density. While the dynamics is faster compared to the cold ion limit, as shown for the case of $\tau_{\text{i}}=1$ in Fig.~\ref{fig:chargedensity}(ii), the charge density helps with smoothing out the fluid densities such that small scales do not develop to the same extent. At finite densities, the corrections to the fields are therefore suppressed to a much larger degree than in the corresponding quasi-neutral limit $\lambda = 0$ where negative densities are encountered. However, the \second-order correction to the charge density is still substantial. In all cases, it therefore seems more appropriate to use the perturbation equations given in App.~\ref{sec:alternativeperb} in the dilute regime. Here, the charge separation is appropriately assumed to be large and can appear at the lowest-order.

\section{Discussion}

We have considered a charged non-relativistic fluid system governed by the Lorentz force where in particular, the electric field were treated self-consistently through the back-reaction on the particle dynamics. The system was considered in a perturbative setup suitable for studying the long wavelength and low frequency dynamics of the system. Expanding the dynamical fields in a formal series expansion a set of perturbative equations for the \first-order corrected fields were obtained. We showed that the perturbation equations preserved the conserved charges of the full underlying system at each order of the expansion. Furthermore, we showed that a non-trivial quasi-neutral limit was possible only because it does not allow charge separating terms at the lowest order. In particular, this excluded the diamagnetic-drift velocity to appear at the lowest order in the expansion. Perturbative schemes that allows the diamagnetic-drift velocity at the lowest order were given in App.~\ref{sec:alternativeperb}. These schemes do not allow a non-trivial quasi-neutral limit as the polarization-drift velocity does not appear at the same order and thus does not introduce time dependence in the electrical current. Finally, the perturbative setup considered in the main text also has a natural generalization to multiple species.

Subsequently, using numerical simulations giving explicit access to the first-order corrections to the dynamical field variables, we have provided a quantitative analysis showing that the perturbation equations are indeed valid at long wavelengths and low frequencies. In particular, we have considered three different parameter regimes from which we summaries: (i) For zero ion temperature, the lowest-order approximation can provide a fairly converged solution to the dynamics at sufficient large scales (if the corrected velocity field is used). (ii) For non-zero ion temperature, the dynamics generally happens on faster time scales. When the ion temperature is non-zero, but sufficiently low the \first-order corrected fields provide a relatively converged approximation, while above a certain temperature higher-order corrections are required in order to resolve the faster dynamics. (iii) For finite densities, the Poisson equation gives rise to a finite charge density (i.e. a split between the density for the species). In this case the dynamics happens on a slower time scale. However, the small scale structures in the electric potential becomes increasingly more important. In this regime, the dilute perturbation equations given in App.~\ref{sec:alternativeperb} becomes applicable. Furthermore, we have provided some approximative scaling transformations that relates the solution between different set of parameters in the cases where the system is density scale invariant (i.e. in the quasi-neutral limit). We have verified that the scaling transformations are in fairly good agreement with the numerical simulations. In conclusion, we have addressed all the points (i)-(iv) listed in the introduction.

\vspace{0.2cm}

This work establishes the very first step in developing perturbative drift-fluid equations where all dynamical fields are under full perturbative control. There are therefore several natural extensions: Firstly, it would be interesting to lift the assumption of a vanishing parallel velocity component $v_{a,\parallel}$ generalizing the equations to a full (3+1)-dimensional system. In particular, this poses two challenges. Since the parallel component can not be treated in the same singular perturbation as the perpendicular components of the velocity, it must be treated dynamically. Its evolution can therefore happen on a different time scale. Also, the relative magnitudes between the components of the electrical current could potentially be large, thus posing a challenge if the system is required to be neutral at the \zeroth order. Secondly, it would be desirable to lift the isothermal assumption on the fluid thereby making the internal energy dynamical. Besides adding additional degrees of freedom to the system, this should in principle a straight-forward extension. Finally, it would be interesting to include dissipative corrections, like viscosity or other resistive terms and furthermore check how these modify the conserved currents. 

Another direction that hold interest is to include a self-consistent magnetic back-reaction on the particle dynamics. This can be done by either including higher-order corrections from the non-relativistic expansion of the Maxwell and fluid equations, or by using the magnetic limit of the Maxwell equations (which is the regime of magnetohydrodynamics). Also, in scrape-off layer plasmas there are often regions, e.g. near boundaries where the plasma density is much smaller. It would be interesting if one could solve the dynamics in these regions using the dilute perturbation equations presented in the App.~\ref{sec:alternativeperb} and feed the solution (self-consistently) as boundary conditions to the presented perturbation equations in Sec.~\ref{sec:scheme2} and vice versa.

\paragraph{Acknowledgements}

We would like to thank Jens Juul Rasmussen for many helpful discussions and suggestions for improving the draft. Furthermore, we would also like to thank Aslak Sindbjerg Poulsen for comments on the earlier drafts.

\newpage
\appendix

\section{Non-relativistic limit and Galilean invariance} \label{sec:nonrel}

The system given by Eq.~\eqref{eq:fluidsystem}, is the non-relativistic limit of a \textit{forced} perfect relativistic fluid with equation of motion $\nabla \cdot T = F \cdot J$. Here, $T$ is the relativistic energy-momentum tensor for a perfect isotropic fluid, $J$ is the relativistic four-current density, and $F$ is the fundamental electromagnetic two-tensor. In the non-relativistic limit the left-hand side reduces to the Euler equations by projection while the right-hand side constitutes the Lorentz force. The relativistic equations are manifestly invariant under Lorentz transformations whereas in the non-relativistic limit the conservation equations together with the Lorentz force (no dependence on acceleration) are invariant under Galilean transformations.

The fluid equations are coupled to the electromagnetic field equations through the Lorentz force. We therefore need a non-relativistic limit of the Maxwell field equations.
Electromagnetism is a Lorentz invariant theory.\footnote{Actually, the Bianchi identities are invariant under Galilean transformations, but the other pair of the Maxwell equations are not.}
However, there exists at least two well-known non-relativistic limits of the Maxwell equations; the electric limit $\mathbf{E} \gg c\mathbf{B}$ and the magnetic limit $\mathbf{E} \ll c\mathbf{B}$ \cite{Manfredi}. We shall be concerned with the electric limit. It is obtained by expanding in the reference velocity $\bar{v}$ normalized to the speed of light $c$, i.e. $\beta = \frac{\bar{v}}{c}$, in the same way the fluid equations are obtained. Introducing the reference length $\ell$ and frequency $\omega$ together with $\bar{E}, \bar{B}, \bar{\rho}$ and $\bar{J}$, satisfying $\bar{v} = \omega \ell$, $\bar{E} = c \bar{B}$ and $\bar{J} = \bar{v} \bar{\rho}$, the dimensionless form of the Maxwell equations is
\begin{align} \label{eq:maxwell}
\begin{split}
\lambda \nabla \cdot \mathbf{E} = \rho ~, \quad
\nabla \cdot \mathbf{B} = 0 ~, \quad
\nabla \times \mathbf{E} = - \beta \partial_t \mathbf{B} ~, \quad
\nabla \times \mathbf{B} = \frac{\beta}{\lambda} \mathbf{J} + \beta \partial_t \mathbf{E} ~.
\end{split}
\end{align}
There are two dimensionless parameters, $\beta$ and $\lambda = \left(\frac{\lambda_D}{\ell}\right)^2$, where $\lambda_D$ is the Debye length (using $\bar{\phi} = \bar{E}\ell = \frac{ k_{\text{B}} \bar{T}}{q}$ with the electric charge $q$ and temperature $\bar{T}$). Now assume that $\beta \ll 1$ and expand the electromagnetic fields in a regular power series $\mathbf{E} = \mathbf{E}^{(0)} + \beta \mathbf{E}^{(1)} + ...$ and $\mathbf{B} = \mathbf{B}^{(0)} + \beta \mathbf{B}^{(1)} + ...$. Here, all coefficients e.g. $\mathbf{E}^{(k)}$ and $\mathbf{B}^{(k)}$ are fields of order unity. This gives to first order
\begin{align} \label{eq:emlimit}
\begin{split}
\lambda \nabla \cdot \mathbf{E}^{(0)} &= \rho ~, \quad \nabla \times \mathbf{E}^{(0)} = \nabla \cdot \mathbf{B}^{(0)} = \nabla \times \mathbf{B}^{(0)} = 0 ~, \\
\nabla \times \mathbf{B}^{(1)} &= \frac{\mathbf{J}}{\lambda} + \partial_t \mathbf{E}^{(0)} ~, \quad \lambda \nabla \cdot \mathbf{E}^{(1)} = \nabla \times \mathbf{E}^{(1)} = \nabla \cdot \mathbf{B}^{(1)} = 0 ~,
\end{split}
\end{align}
which implies that $\mathbf{B}^{(0)} = \mathbf{E}^{(1)} = 0$, if the fields vanish at infinity. Equivalently by introducing the vector potential $\mathbf{B}^{(1)} = \nabla \times \mathbf{A}^{(1)}$ and electric potential $\mathbf{E}^{(0)} = - \nabla \phi^{(0)}$, we can write in the Lorentz gauge
\begin{align} \label{eq:NRelectriclimit}
\begin{split}
 \lambda \nabla^2 \phi^{(0)} = - \rho ~, \quad \lambda \nabla^2 \mathbf{A}^{(1)} = - \mathbf{J} ~.
\end{split}
\end{align}
The electric and magnetic fields thus appear at different orders of $\beta$ and one can check that indeed $\mathbf{E} \gg c\mathbf{B}$. In this limit, the Lorentz force is solely electric as the first order quantities only forms a second-order magnetic correction. The force per unit volume is
\begin{equation}
	\mathbf{F} = \rho \mathbf{E}^{(0)} + \mathcal{O}(\beta)^2 ~.
\end{equation}
The particle dynamics is thus only governed by an electric force.\footnote{Note that the Lorentz force appearing in Eq.~\eqref{eq:fluidsystem} is constituted by $\mathbf{E}^{(0)}$ and  an \textit{external} background field.} So even though the magnetic field exist in the limit it does not contribute to the particle dynamics. The equation for the vector potential given in Eq.~\eqref{eq:NRelectriclimit} can therefore be left out of the dynamical consideration. Dropping the order indices in Eq.~\eqref{eq:NRelectriclimit} we are thus left with Eq.~\eqref{eq:emsystem}. If one is interested in a self-consistent back-reaction on the particle dynamics due to the magnetic field, it is necessary to include higher-order relativistic corrections both in the Maxwell equations as well as the fluid equations. The conservation of charge current density can be obtained by taking the divergence of Eq.~\eqref{eq:emlimit}, $\partial_t \rho + \nabla \cdot \mathbf{J} = 0$.

Finally, it should be checked that the field equations are invariant under Galliean transformations. Considering a general Lorentz transformation and expanding in $\beta$ one finds up to second order the Galliean transformation $\{t' = t, \mathbf{x}' = \mathbf{x} + \mathbf{v} t\}$ such that $\nabla' = \nabla$ and $\partial_t' = \partial_t - \mathbf{v} \cdot \nabla$ together with
\begin{align}
\begin{split}
\mathbf{E}'^{(0)} = \mathbf{E}^{(0)} ~, \quad
\mathbf{B}'^{(1)} = \mathbf{B}^{(1)} - \mathbf{v} \times \mathbf{E}^{(0)} ~, \quad
\mathbf{J}' = \mathbf{J} - \mathbf{v} \rho ~, \quad \rho' = \rho ~,
\end{split}
\end{align}
leaves the equations of motion invariant. We have thus obtained a consistent non-relativistic limit $\beta~\ll~1$ of the Maxwell equations in combination with the non-relativistic fluid equations given by Eq.~\eqref{eq:fluidsystem} and \eqref{eq:emsystem}.

\section{Conserved charges} \label{sec:charges}

In this appendix we consider the conserved quantities associated with the fluid system given by Eq.~\eqref{eq:normalized}. The conservation equation for mass (and charge) is obtained by transforming the continuity equation into an integral equation and using the divergence theorem,
\begin{equation}
\partial_t \int_V \td V n_a + \int_{\partial V} \td \mathbf{A} \cdot \mathbf{v}_a n_a = 0 ~. 
\end{equation}
The integral equation states that the variation of mass (or charge) of a single species in the volume $V$ is given by the flux through the boundary. In a similar way, the total energy can be obtained from the energy conservation equation
\begin{equation} \label{eq:energyconservation}
	\partial_t \left( \varepsilon_a + \frac{\mu_a}{2} n_a \mathbf{v}_a^2 \right) + \nabla \cdot \left( \varepsilon_a + p_a +  \frac{\mu_a}{2} n_a \mathbf{v}_a^2 \right) =  - \nabla \phi \cdot n_a \mathbf{v}_a ~,
\end{equation}
where $\varepsilon_a$ is the energy density. The total energy can be obtained by adding the energy conservation equations \eqref{eq:energyconservation} for all species and integrate over the volume. It is useful to rewrite the variables by using $p_a = n_a\tau_a$ and the Euler relation $\varepsilon_a + p_a = s_a\tau_a + z_a n_a\phi$, with the entropy density $s_a$.  The total energy is constituted by the kinetic, potential and heat contributions. The kinetic energy contribution is obtained by considering the integral over $\partial_t (n_a \mathbf{v}_a^2)$ using the continuity and momentum equation,
\begin{equation} \label{eq:kinetic}
\frac{\mu_a}{2} \partial_t \int \td V n_a \mathbf{v}_a^2 + \frac{\mu_a}{2} \int \td \mathbf{A} \cdot \mathbf{v}_a n_a \mathbf{v}_a^2 = -  \int \td V \mathbf{v}_a \cdot ( \tau_a \nabla n_a + z_a n_a \nabla \phi) ~. 
\end{equation}
The potential energy in the electric potential can be found from Poisson's equation and the continuity equations for positively and negatively charged particles,
\begin{equation} \label{eq:potenergy}
\frac{\lambda}{2} \partial_t \int \td V (\nabla \phi)^2 + \int \td \mathbf{A} \cdot \left[ - \frac{\lambda}{2} \partial_t (\nabla \phi) + (\mathbf{v}_{\text{i}} n_{\text{i}} - \mathbf{v}_{\text{e}} n_{\text{e}} ) \right] \phi = \int \td V \nabla \phi \cdot (\mathbf{v}_{\text{i}} n_{\text{i}} - \mathbf{v}_{\text{e}} n_{\text{e}} ) ~. 
\end{equation}
Finally, the entropy density in a system in contact with a heat reservoir is $s_a = n_a(\log n_a + 1)$. The heat contribution from a single species is therefore found by multiplying the continuity equation by $\tau_a (\log n_a + 1)$ and integrating over the total volume
\begin{equation} \label{eq:heat}
\tau_a \partial_t \int \td V n_a \log n_a + \tau_a \int \td \mathbf{A} \cdot \mathbf{v}_a  n_a (1 + \log n_a ) = \tau_a \int \td V \mathbf{v}_a \cdot \nabla n_a  ~. 
\end{equation}
Integrating Eq.~\eqref{eq:energyconservation} and using Eq.~\eqref{eq:kinetic} and Eq.~\eqref{eq:heat} for all species together with Eq.~\eqref{eq:potenergy}, the conservation of energy is
\begin{align} \label{eq:totalenergy}
\begin{split}
\partial_t \int \td V &\left[ \frac{\lambda}{2} (\nabla \phi)^2 +
\sum_{a} 
\tau_{a} n_{a} \log n_{a} + 
\frac{\mu_{a}}{2} n_{a} \mathbf{v}_{a}^2
\right] + \\
 \int \td \mathbf{A} \cdot &\left[
- \frac{\lambda}{2} \partial_t (\nabla \phi) \phi +
\sum_{a} 
\mathbf{v}_{a} n_{a} \left( \tau_{a} (1 + \log n_{a}) +\frac{\mu_{a}}{2} \mathbf{v}_{a}^2 + z_a \phi \right)
\right] = 0~. 
\end{split}
\end{align}
If the dynamical fields vanish at the boundary the total energy will be a constant of motion.

\section{Alternative perturbative setups} \label{sec:alternativeperb}

In this section we present two alternative perturbative setups that both can handle charge separation at the lowest order. However, we show that it is not possible to have a consistent quasi-neutral limit in such setups. Consequently, one must make use of the Poisson equation and the setups therefore only makes sense for low fluid densities.

\subsubsection*{Perturbative setup for low densities}

Small scale structures of the electric potential becomes increasingly important as the fluid becomes more dilute $\lambda\sim\mathcal{O}(1)$. In such cases, a more suitable setup is obtained by choosing $\omega = \omega_{\text{c,i}}$ and scaling all the fields $n_a \rightarrow \varepsilon n_a$, $\phi \rightarrow \varepsilon \phi$, $\mathbf{v}_a \rightarrow \varepsilon \mathbf{v}_a$ together with $t \rightarrow \varepsilon^{-1}t$. This way, the spatial structure has more significance while only the slow dynamics is considered. In other words, the electric field is obtained from Poisson equation and we have
\begin{align} 
\begin{split}
	\lambda\nabla^2 \phi &= - \rho ~, \quad
	\partial_t n_a + \nabla \cdot ( n_a \mathbf{v}_a ) = 0 ~, \\
	\varepsilon \mu_a n_a \text{d}_t \mathbf{v}_a &= - \tau_a\nabla n_a + z_a n_a ( - \nabla \phi + \mathbf{v}_a \times \mathbf{B} ) ~.
\end{split}
\end{align}
Charge separating terms are now allowed at \zeroth order, i.e. the diamagnetic-drift velocity. In fact, if it is pushed to \first order, the \zeroth order density will simply work as a static background for the higher orders. The set of equations at order $k$ has the form
\begin{align} \label{eq:scheme1zeroth}
\lambda\nabla^2 \phi^{(k)} = - \rho^{(k)} ~, \quad
\partial_t n^{(k)}_a + \nabla \cdot \mathbf{\Gamma}^{(k)}_a = 0 ~.
\end{align}
We therefore do not have a neutral system at the \zeroth order of the expansion.
The \zeroth order is given by
\begin{align}
\mathbf{\Gamma}^{(0)}_a = n^{(0)}_a \mathbf{v}^{(0)}_{a} ~, \quad
\mathbf{v}^{(0)}_{a} = \mathbf{v}^{(0)}_{\text{E}} + \mathbf{v}^{(0)}_{\text{D}a}~, \quad
\mathbf{u}_{\text{D}a}^{(0)} = \tau_{a} \frac{\mathbf{b} \times \nabla \log n^{(0)}_a}{B} ~,
\end{align}
while the \first order is given by
\begin{align}
\mathbf{\Gamma}^{(1)}_a = n^{(0)}_a ( \mathbf{v}^{(1)}_{\text{E}} + \mathbf{v}^{(1)}_{\text{D}a} + \mathbf{v}^{(0)}_{\text{P}a} ) + n^{(1)}_a \mathbf{v}^{(0)}_{\text{E}} ~, \quad
\mathbf{u}_{\text{P}a}^{(0)} = z_a \mu_a \frac{\mathbf{b} \times \td_t \mathbf{v}^{(0)}_{a} }{B} ~.
\end{align}
The polarization-drift velocity is obtained by taking time derivatives of Eq.~\eqref{eq:scheme1zeroth} and solving for the time-derivative of the potential at a given order. One significant difference between the this scheme and the one presented in Sec.~\ref{sec:scheme2} is the clear separation between the different orders. Whereas it was necessary to go to the next order to close the current set of equations in Sec.~\ref{sec:scheme2}, the equations closes at each order in the current scheme. Also note that the equations are fully capable of handling multiple species.

\subsubsection*{Perturbative setup with suppressed diamagnetic advection} \label{sec:Gsplit}

A posteriori the singular drift expansion, one can argue that while the diamagnetic-drift velocity is a large contribution to the velocity it only gives a small contribution to the density evolution (as it does not advect). This is particularly the case if the gradients of the magnetic field are small. One proposal is therefore to let the $\text{E}{\times}\text{B}$-drift velocity perform the advection and push the contribution from the diamagnetic-drift velocity to the density flux to next order. We shall show that this is indeed possible, but the scheme does not allow for quasi-neutrality, since in this limit it correspond to a spatial constraint. To implement this idea we split the velocity field such that $\mathbf{v}_a \equiv \mathbf{v}_{\text{E}} + \mathbf{v}_{\text{R}a}$ where $\mathbf{v}_{\text{E}}$ accounts for the $\text{E}{\times}\text{B}$-drift velocity and $\mathbf{v}_{\text{R}a}$ contains the rest. We can then consider the perturbation problem
\begin{align} 
\begin{split}
\varepsilon\lambda\nabla^2 \phi &= - \rho ~, \quad
\partial_t n_a + \nabla \cdot ( n_a \mathbf{v}_{\text{E}} ) + \varepsilon \nabla \cdot \left( n_a \mathbf{v}_{\text{R}a} \right) = 0 ~, \\
\varepsilon \mu_a n_a \text{d}_t \mathbf{v}_a &= - \tau_a\nabla n_a + z_a n_a \left( - \nabla \phi + \mathbf{v}_a \times \mathbf{B} \right) ~.
\end{split}
\end{align}
The two first terms of the perpendicular velocities are then
\begin{align}
\mathbf{v}^{(0)}_{a,\perp} = \mathbf{v}^{(0)}_{\text{E}} + \mathbf{v}^{(0)}_{\text{D}a} ~,\quad
\mathbf{v}^{(1)}_{a,\perp} = \mathbf{v}^{(1)}_{\text{E}} + \mathbf{v}^{(1)}_{\text{D}a} + \mathbf{v}^{(0)}_{\text{P}a} ~,
\end{align}
and we identify
\begin{align}
\mathbf{v}^{(0)}_{\text{R}a} = \mathbf{v}^{(0)}_{\text{D}a} ~, \quad
\mathbf{v}^{(1)}_{\text{R}a} = \mathbf{v}^{(1)}_{\text{D}a} + \mathbf{v}^{(0)}_{\text{P}a} ~.
\end{align}
At \zeroth order we have a neutral system, i.e. $\rho^{(0)} = 0$, we can thus define a single density $n^{(0)} = n^{(0)}_{\text{i}} = n^{(0)}_{\text{e}}$ satisfying the \zeroth order continuity equation given by Eq.~\eqref{eq:continuity0}. The \first-order continuity equation and charge-current conservation equation are
\begin{align}
\partial_t n^{(1)}_a &+ \nabla \cdot \left( n^{(0)} \mathbf{v}^{(1)}_{\text{E}} + n^{(1)}_a \mathbf{v}^{(0)}_{\text{E}} + n^{(0)} \mathbf{v}^{(0)}_{\text{D}a} \right) = 0 ~. \\
\partial_t \rho^{(1)} &+ \nabla \cdot \left( \rho^{(1)} \mathbf{v}^{(0)}_{\text{E}} + n^{(0)} \mathbf{v}^{(0)}_{\text{D}} \right) = 0 ~. \label{eq:Gsplitcharge}
\end{align}
where $\mathbf{v}_{\text{D}}^{(0)} = \mathbf{v}_{\text{Di}}^{(0)} - \mathbf{v}_{\text{De}}^{(0)}$.
However, in the quasi-neutral limit $\lambda = 0$, Eq.~\eqref{eq:Gsplitcharge} reduces to the spatial constraint equation $\nabla \cdot ( n^{(0)} \mathbf{v}^{(0)}_{\text{D}} ) = 0$, the simple reason being that the polarization-drift velocity enters at the \second order. This constraint can only be satisfied if either the gradient of the magnetic field vanish or if there are no \zeroth-order pressure gradients (in which case the \zeroth-order diamagnetic-drift velocity vanish). Furthermore, in this limit there is no dynamical equation for the \zeroth-order electric potential which must vanish.

\addcontentsline{toc}{section}{References}
\AtNextBibliography{\scriptsize}
\printbibliography

\end{document}

%% file: coldiontable.tex
\begin{table}[H]
\centering
\ra{1.3}
\begin{tabular}{@{}lll@{}}
\toprule
Field & Time & Value \\
\midrule
$C_2^{(1)}(\phi)$ &
\(\displaystyle \sigma^{\frac{1}{2}} \tilde{A}^{-1} \) &
\(\displaystyle \sigma^{-\frac{1}{2}} \tilde{A}^{-1} \) \\[1mm]
$C_{2}^{(1)}(n)$ & 
\(\displaystyle \sigma^{\frac{1}{2}} \tilde{A}^{-1} \) & 
\(\displaystyle \sigma^{-\frac{1}{2}} A^{\frac{1}{2}} \) \\[1mm]
$C_2^{(1)}(\mathbf{v}_{\text{e}})$, $C_2^{(1)}(\mathbf{\tilde{v}}_{\text{e}})$
& 
\(\displaystyle \sigma^{\frac{1}{2}} \tilde{A}^{-\frac{1}{2}} \) &
\(\displaystyle \sigma^{-\frac{3}{2}} \tilde{A}^{\frac{1}{2}} \) \\[1mm]
$C_2^{(1)}(\mathbf{v}_{\text{i}})$, $C_2^{(1)}(\mathbf{\tilde{v}}_{\text{i}})$
& 
\(\displaystyle \sigma^{\frac{1}{2}} \) &
\(\displaystyle \sigma^{-\frac{1}{2}} \) \\[1mm]
\bottomrule
\end{tabular}
\caption{Scaling transformations of the operator $C$ applied on the fields in the quasi-neutral ($\lambda=0$) cold ion ($\tau_{\text{i}} = 0$) limit with respect to $\{A, \sigma\}$. Here $\tilde{A} \equiv \frac{A}{1+A}$. The scaling relations hold to good approximation with respect to $\sigma$ while the relations for the amplitude $A$ only holds for early times as shown in Fig.~\ref{fig:sigmascaling}. Note that for the density, the value scales with $A$ and not $\tilde{A}$.} 
\label{tab:coldion}
\end{table}

%% file: coldevolution.tex
\begin{figure}[H]
\centering
\includegraphics[width=0.32\textwidth]{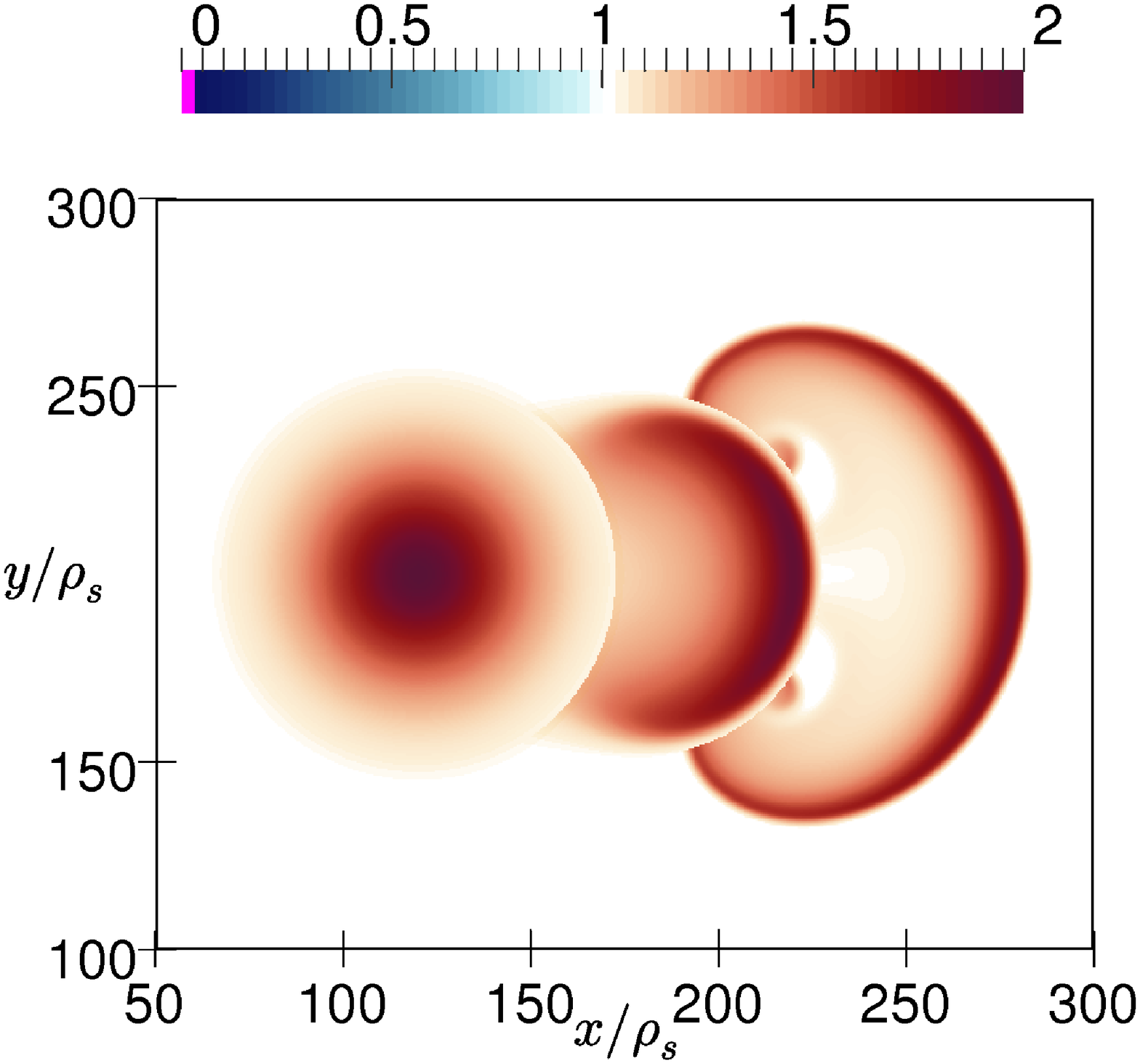}
\includegraphics[width=0.32\textwidth]{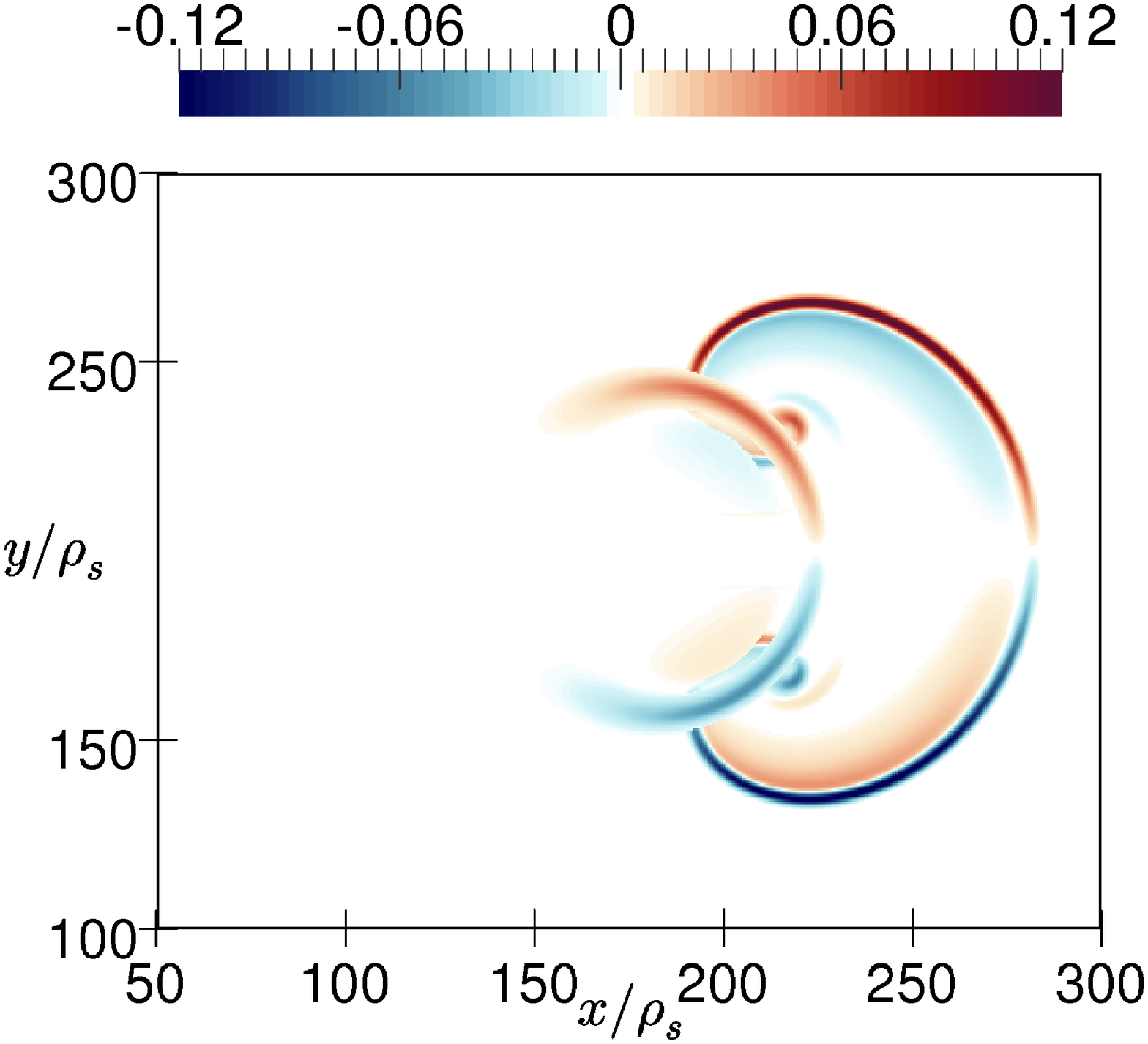}
\includegraphics[width=0.32\textwidth]{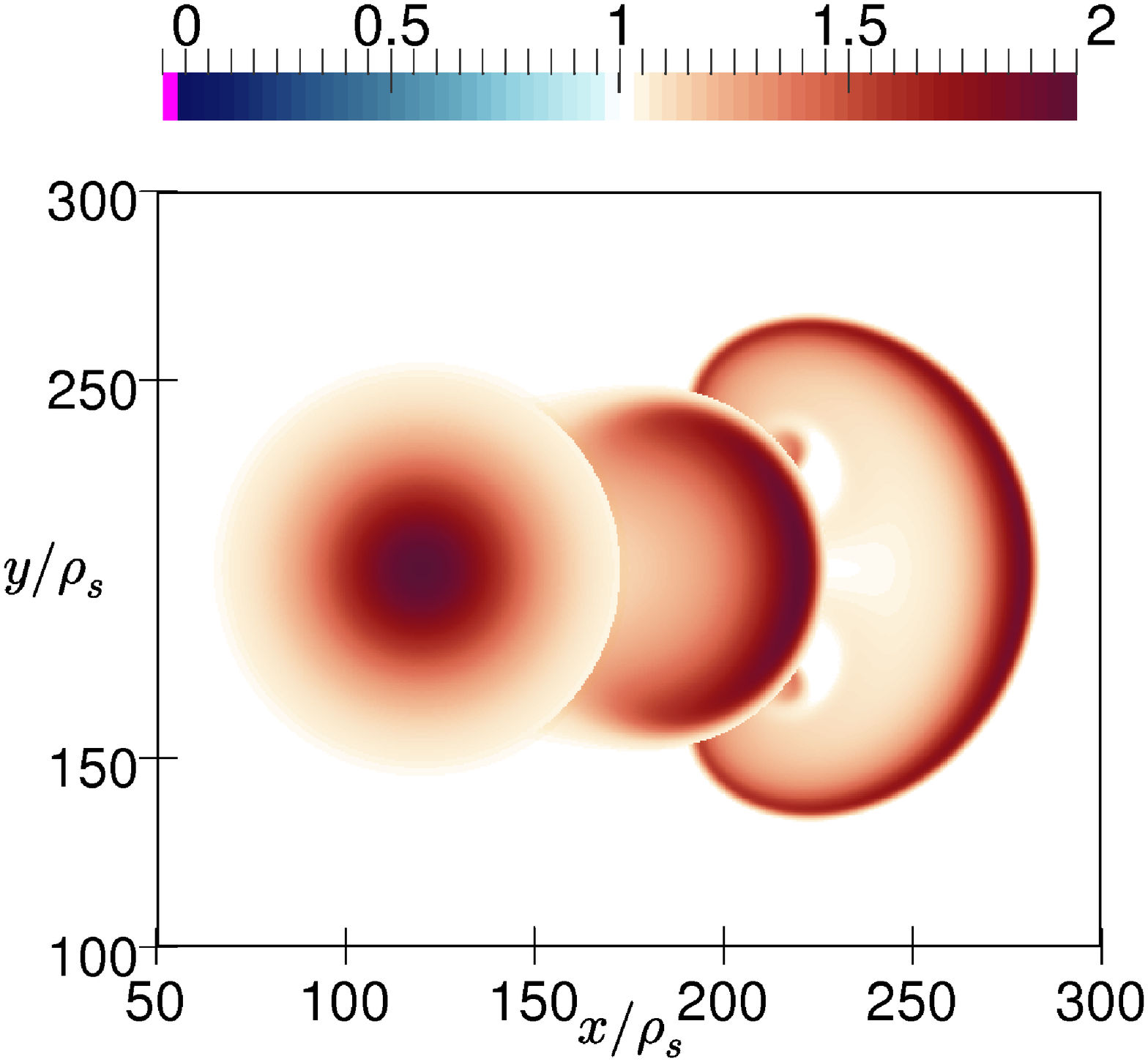}
\caption{The density field in the quasi-neutral limit $\lambda=0$ and cold ion limit $\tau_{\text{i}} = 0$ for the parameters $A=1$ and $\sigma=20$. From left to right, the field $n^{(0)}$, $n^{(1)}$, and the \first-order approximation of the density $n \approx n^{(0)} + n^{(1)}$ are shown (overlapped) at three different times $\{0, 2450, 3770\}\omega_{\text{c,i}}$.}
\label{fig:evolution0}
\end{figure}

%% file: sigmascaling.tex
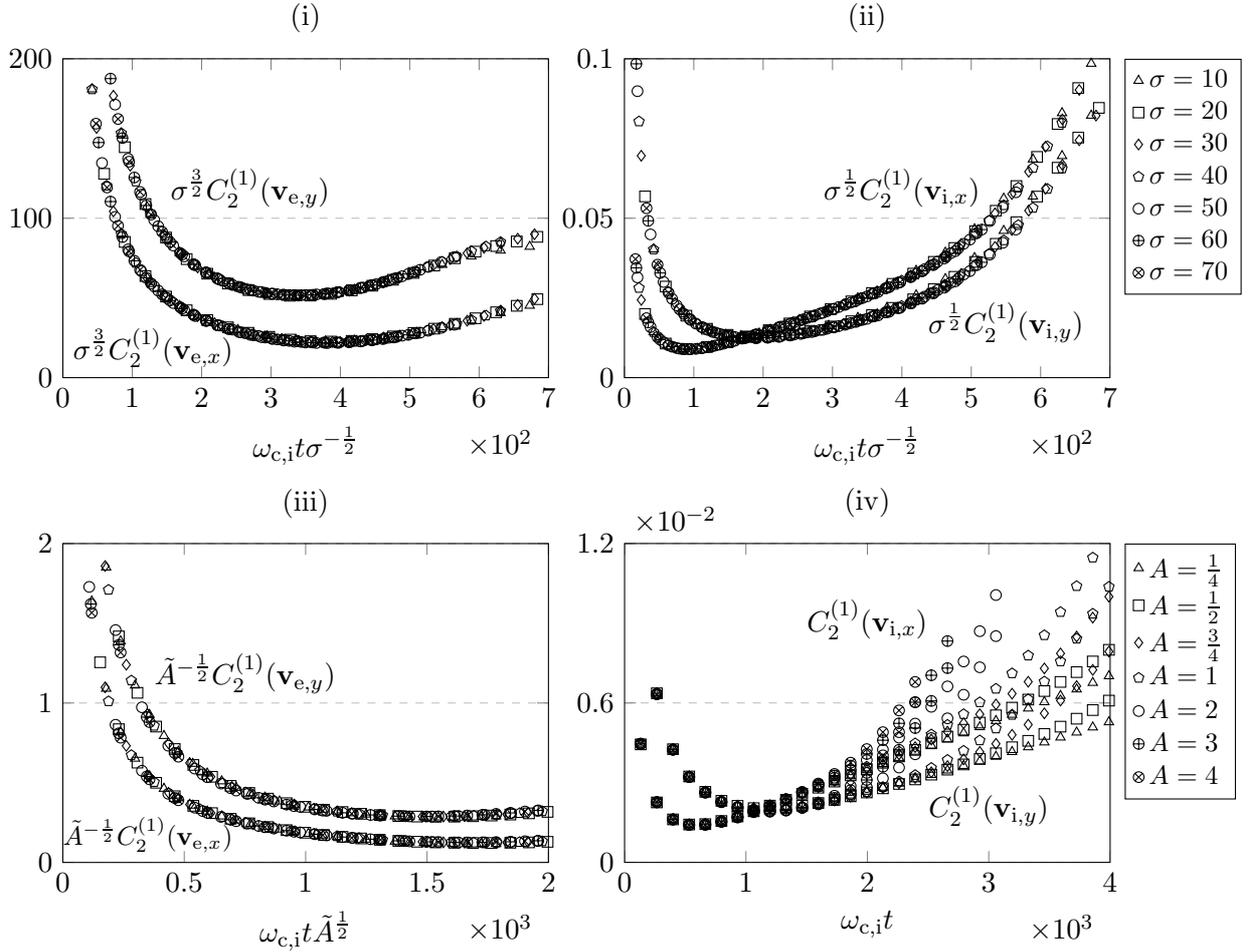
\begin{figure}[H]
\centering
\begin{tikzpicture}
\begin{groupplot}[
group style={rows=2, columns=2, vertical sep=2.2cm},
width=0.48\textwidth,
height=0.35\textwidth,
legend cell align={left},
]

\nextgroupplot[
title={(i)},
tick scale binop=\times,
xlabel={$\omega_{\text{c,i}} t \sigma^{-\frac{1}{2}}$},
xmin=0, xmax=700,
ymin=0, ymax=200,
xtick={0, 100, 200, 300, 400, 500, 600, 700},
scaled x ticks={base 10:-2},
ytick={0, 100, 200},
/pgf/number format/.cd, 1000 sep={},
ymajorgrids=true,
grid style=dashed,
]
\node [anchor=south, color=black] at (axis cs:130,0) {$\sigma^{\frac{3}{2}}C_2^{(1)}(\mathbf{v}_{\text{e},x})$};
\node [anchor=south, color=black] at (axis cs:270,100) {$\sigma^{\frac{3}{2}}C_2^{(1)}(\mathbf{v}_{\text{e},y})$};
\addplot[style_1] table [x={time10}, y={v1ex_v0ex_10}]{figures/scalingve.txt};
\addplot[style_2] table [x={time20}, y={v1ex_v0ex_20}]{figures/scalingve.txt};
\addplot[style_3] table [x={time30}, y={v1ex_v0ex_30}]{figures/scalingve.txt};
\addplot[style_4] table [x={time40}, y={v1ex_v0ex_40}]{figures/scalingve.txt};
\addplot[style_5] table [x={time50}, y={v1ex_v0ex_50}]{figures/scalingve.txt};
\addplot[style_6] table [x={time60}, y={v1ex_v0ex_60}]{figures/scalingve.txt};
\addplot[style_7] table [x={time70}, y={v1ex_v0ex_70}]{figures/scalingve.txt};

\addplot[style_1, forget plot] table [x={time10}, y={v1ey_v0ey_10}]{figures/scalingve.txt};
\addplot[style_2, forget plot] table [x={time20}, y={v1ey_v0ey_20}]{figures/scalingve.txt};
\addplot[style_3, forget plot] table [x={time30}, y={v1ey_v0ey_30}]{figures/scalingve.txt};
\addplot[style_4, forget plot] table [x={time40}, y={v1ey_v0ey_40}]{figures/scalingve.txt};
\addplot[style_5, forget plot] table [x={time50}, y={v1ey_v0ey_50}]{figures/scalingve.txt};
\addplot[style_6, forget plot] table [x={time60}, y={v1ey_v0ey_60}]{figures/scalingve.txt};
\addplot[style_7, forget plot] table [x={time70}, y={v1ey_v0ey_70}]{figures/scalingve.txt};


\nextgroupplot[
title={(ii)},
tick scale binop=\times,
xlabel={$\omega_{\text{c,i}} t \sigma^{-\frac{1}{2}}$},
xmin=0, xmax=700,
ymin=0, ymax=0.1,
xtick={0, 100, 200, 300, 400, 500, 600, 700},
scaled x ticks={base 10:-2},
ytick={0, 0.05, 0.1},
y tick label style={/pgf/number format/fixed},
/pgf/number format/.cd, 1000 sep={},
ymajorgrids=true,
grid style=dashed,
legend pos=outer north east,
legend style={font=\small},
]

\node [anchor=south, color=black] at (axis cs:400,0.05) {$\sigma^{\frac{1}{2}}C_2^{(1)}(\mathbf{v}_{\text{i},x})$};
\node [anchor=south, color=black] at (axis cs:550,0.008) {$\sigma^{\frac{1}{2}}C_2^{(1)}(\mathbf{v}_{\text{i},y})$};

\addplot[style_1] table [x={time10}, y={v1ix_v0ix_10}]{figures/scalingvi.txt};
\addplot[style_2] table [x={time20}, y={v1ix_v0ix_20}]{figures/scalingvi.txt};
\addplot[style_3] table [x={time30}, y={v1ix_v0ix_30}]{figures/scalingvi.txt};
\addplot[style_4] table [x={time40}, y={v1ix_v0ix_40}]{figures/scalingvi.txt};
\addplot[style_5] table [x={time50}, y={v1ix_v0ix_50}]{figures/scalingvi.txt};
\addplot[style_6] table [x={time60}, y={v1ix_v0ix_60}]{figures/scalingvi.txt};
\addplot[style_7] table [x={time70}, y={v1ix_v0ix_70}]{figures/scalingvi.txt};

\addplot[style_1, forget plot] table [x={time10}, y={v1iy_v0iy_10}]{figures/scalingvi.txt};
\addplot[style_2, forget plot] table [x={time20}, y={v1iy_v0iy_20}]{figures/scalingvi.txt};
\addplot[style_3, forget plot] table [x={time30}, y={v1iy_v0iy_30}]{figures/scalingvi.txt};
\addplot[style_4, forget plot] table [x={time40}, y={v1iy_v0iy_40}]{figures/scalingvi.txt};
\addplot[style_5, forget plot] table [x={time50}, y={v1iy_v0iy_50}]{figures/scalingvi.txt};
\addplot[style_6, forget plot] table [x={time60}, y={v1iy_v0iy_60}]{figures/scalingvi.txt};
\addplot[style_7, forget plot] table [x={time70}, y={v1iy_v0iy_70}]{figures/scalingvi.txt};


\addlegendentry{$\sigma=10$} 
\addlegendentry{$\sigma=20$}
\addlegendentry{$\sigma=30$}
\addlegendentry{$\sigma=40$}
\addlegendentry{$\sigma=50$}
\addlegendentry{$\sigma=60$}
\addlegendentry{$\sigma=70$}

\nextgroupplot[
title={(iii)},
tick scale binop=\times,
xlabel={$\omega_{\text{c,i}} t \tilde{A}^{\frac{1}{2}}$},
xmin=0, xmax=2000,
ymin=0, ymax=2,
xtick={0, 500, 1000, 1500, 2000},
scaled x ticks={base 10:-3},
ytick={0, 1, 2},
/pgf/number format/.cd, 1000 sep={},
ymajorgrids=true,
grid style=dashed,
]
\node [anchor=south, color=black] at (axis cs:350,0) {\small $\tilde{A}^{-\frac{1}{2}}C_2^{(1)}(\mathbf{v}_{\text{e},x})$};
\node [anchor=south, color=black] at (axis cs:750,1) {$\tilde{A}^{-\frac{1}{2}}C_2^{(1)}(\mathbf{v}_{\text{e},y})$};

\addplot[style_1] table [x={time_025}, y={v1ex_v0ex_025}]{figures/scalingampve.txt};
\addplot[style_2] table [x={time_05}, y={v1ex_v0ex_05}]{figures/scalingampve.txt};
\addplot[style_3] table [x={time_075}, y={v1ex_v0ex_075}]{figures/scalingampve.txt};
\addplot[style_4] table [x={time_1}, y={v1ex_v0ex_1}]{figures/scalingampve.txt};
\addplot[style_5] table [x={time_2}, y={v1ex_v0ex_2}]{figures/scalingampve.txt};
\addplot[style_6] table [x={time_3}, y={v1ex_v0ex_3}]{figures/scalingampve.txt};
\addplot[style_7] table [x={time_4}, y={v1ex_v0ex_4}]{figures/scalingampve.txt};

\addplot[style_1, forget plot] table [x={time_025}, y={v1ey_v0ey_025}]{figures/scalingampve.txt};
\addplot[style_2, forget plot] table [x={time_05}, y={v1ey_v0ey_05}]{figures/scalingampve.txt};
\addplot[style_3, forget plot] table [x={time_075}, y={v1ey_v0ey_075}]{figures/scalingampve.txt};
\addplot[style_4, forget plot] table [x={time_1}, y={v1ey_v0ey_1}]{figures/scalingampve.txt};
\addplot[style_5, forget plot] table [x={time_2}, y={v1ey_v0ey_2}]{figures/scalingampve.txt};
\addplot[style_6, forget plot] table [x={time_3}, y={v1ey_v0ey_3}]{figures/scalingampve.txt};
\addplot[style_7, forget plot] table [x={time_4}, y={v1ey_v0ey_4}]{figures/scalingampve.txt};

\nextgroupplot[
title={(iv)},
tick scale binop=\times,
xlabel={$\omega_{\text{c,i}} t$},
xmin=0, xmax=4000,
ymin=0, ymax=0.012,
xtick={0, 1000, 2000, 3000, 4000},
scaled x ticks={base 10:-3},
ytick={0, 0.006, 0.012},
/pgf/number format/.cd, 1000 sep={},
ymajorgrids=true,
grid style=dashed,
legend pos=outer north east,
legend style={font=\small},
]
\node [anchor=south, color=black] at (axis cs:2000,0.008) {$C_2^{(1)}(\mathbf{v}_{\text{i},x})$};
\node [anchor=south, color=black] at (axis cs:3000,0.001) {$C_2^{(1)}(\mathbf{v}_{\text{i},y})$};

\addplot[style_1] table [x={time}, y={v1ix_v0ix_025}]{figures/scalingampvi.txt};
\addplot[style_2] table [x={time}, y={v1ix_v0ix_05}]{figures/scalingampvi.txt};
\addplot[style_3] table [x={time}, y={v1ix_v0ix_075}]{figures/scalingampvi.txt};
\addplot[style_4] table [x={time}, y={v1ix_v0ix_1}]{figures/scalingampvi.txt};
\addplot[style_5, restrict x to domain=0:3059] table [x={time}, y={v1ix_v0ix_2}]{figures/scalingampvi.txt};
\addplot[style_6, restrict x to domain=0:2660] table [x={time}, y={v1ix_v0ix_3}]{figures/scalingampvi.txt};
\addplot[style_7, restrict x to domain=0:2394] table [x={time}, y={v1ix_v0ix_4}]{figures/scalingampvi.txt};

\addplot[style_1, forget plot] table [x={time}, y={v1iy_v0iy_025}]{figures/scalingampvi.txt};
\addplot[style_2, forget plot] table [x={time}, y={v1iy_v0iy_05}]{figures/scalingampvi.txt};
\addplot[style_3, forget plot] table [x={time}, y={v1iy_v0iy_075}]{figures/scalingampvi.txt};
\addplot[style_4, forget plot] table [x={time}, y={v1iy_v0iy_1}]{figures/scalingampvi.txt};
\addplot[style_5, forget plot, restrict x to domain=0:3059] table [x={time}, y={v1iy_v0iy_2}]{figures/scalingampvi.txt};
\addplot[style_6, forget plot, restrict x to domain=0:2660] table [x={time}, y={v1iy_v0iy_3}]{figures/scalingampvi.txt};
\addplot[style_7, forget plot, restrict x to domain=0:2394] table [x={time}, y={v1iy_v0iy_4}]{figures/scalingampvi.txt};

\addlegendentry{$A=\frac{1}{4}$} 
\addlegendentry{$A=\frac{1}{2}$}
\addlegendentry{$A=\frac{3}{4}$}
\addlegendentry{$A=1$}
\addlegendentry{$A=2$}
\addlegendentry{$A=3$}
\addlegendentry{$A=4$}

\end{groupplot}
\end{tikzpicture}
\caption{The scaling transformations listed in Tab.~\ref{tab:coldion} for the operator $C$ applied on the component of the electron (left) and ion velocity (right) for (i, ii) fixed $\{\lambda=0,\tau_{\text{i}}=0,A=1\}$ with varying width $\sigma$, and (iii, iv) fixed $\{\lambda=0,\tau_{\text{i}}=0,\sigma=40\}$ with varying amplitude $A$. Here, $\tilde{A} \equiv \frac{A}{1+A}$.}
\label{fig:sigmascaling}
\end{figure}

%% file: cold0.tex
\begin{figure}[H]
\centering
\begin{tikzpicture}
\begin{groupplot}[
group style={rows=1, columns=2, vertical sep=2.0cm},
width=0.5\textwidth,
height=0.35\textwidth,
legend cell align={left},
cycle list name = my BandW,
]

\nextgroupplot[
title={(i)},
tick scale binop=\times,
xlabel={$\omega_{\text{c,i}} t$},
xmin=0, xmax=9000,
ymin=0, ymax=0.5,
xtick={0, 1000, 2000, 3000, 4000, 5000, 6000, 7000, 8000, 9000},
scaled x ticks={base 10:-3},
ytick={0, 0.25, 0.5},
/pgf/number format/.cd, 1000 sep={},
ymajorgrids=true,
grid style=dashed,
legend pos=north east,
legend columns=2,
] 
\addplot table [x={time}, y={v1ex_v0ex}]{figures/sigma50_2.txt};
\addlegendentry{$C_2^{(1)}(\mathbf{v}_{\text{e},x})$} 
\addplot table [x={time}, y={v1ey_v0ey}]{figures/sigma50_2.txt};
\addlegendentry{$C_2^{(1)}(\mathbf{v}_{\text{e},y})$}
\addplot table [x={time}, y={v1ix_v0ix}]{figures/sigma50_2.txt};
\addlegendentry{$C_2^{(1)}(\mathbf{v}_{\text{i},x})$} 
\addplot table [x={time}, y={v1iy_v0iy}]{figures/sigma50_2.txt};
\addlegendentry{$C_2^{(1)}(\mathbf{v}_{\text{i},y})$}

\nextgroupplot[
title={(ii)},
tick scale binop=\times,
xlabel={$\omega_{\text{c,i}} t$},
xmin=0, xmax=9000,
ymin=0, ymax=0.3,
xtick={0, 1000, 2000, 3000, 4000, 5000, 6000, 7000, 8000, 9000},
scaled x ticks={base 10:-3},
ytick={0, 0.1, 0.2, 0.3},
/pgf/number format/.cd, 1000 sep={},
ymajorgrids=true,
grid style=dashed,
legend pos=north west,
] 
\addplot table [x={time}, y={vt2ex_v1ex}]{figures/sigma50.txt};
\addlegendentry{$C_2^{(2)}(\mathbf{v}_{\text{e},x})^{\ast}$} 
\addplot table [x={time}, y={vt2ey_v1ey}]{figures/sigma50.txt};
\addlegendentry{$C_2^{(2)}(\mathbf{v}_{\text{e},y})^{\ast}$}
\addplot table [x={time}, y={vt2ix_v1ix}]{figures/sigma50.txt};
\addlegendentry{$C_2^{(2)}(\mathbf{v}_{\text{i},x})^{\ast}$} 
\addplot table [x={time}, y={vt2iy_v1iy}]{figures/sigma50.txt};
\addlegendentry{$C_2^{(2)}(\mathbf{v}_{\text{i},y})^{\ast}$}

\end{groupplot}
\end{tikzpicture}
\caption{The operator $C$ applied on the electron and ion velocity components. Both figures show the absolute values for the particular simulation parameters $\{\lambda=0, \tau_{\text{i}}=0, A=1, \sigma=50\}$. $^{\ast}$The \second-order correction to the velocity field \eqref{eq:vel012} is approximated by neglecting $\mathbf{v}^{(2)}_{\text{E}}$. }
\label{fig:cold0}
\end{figure}
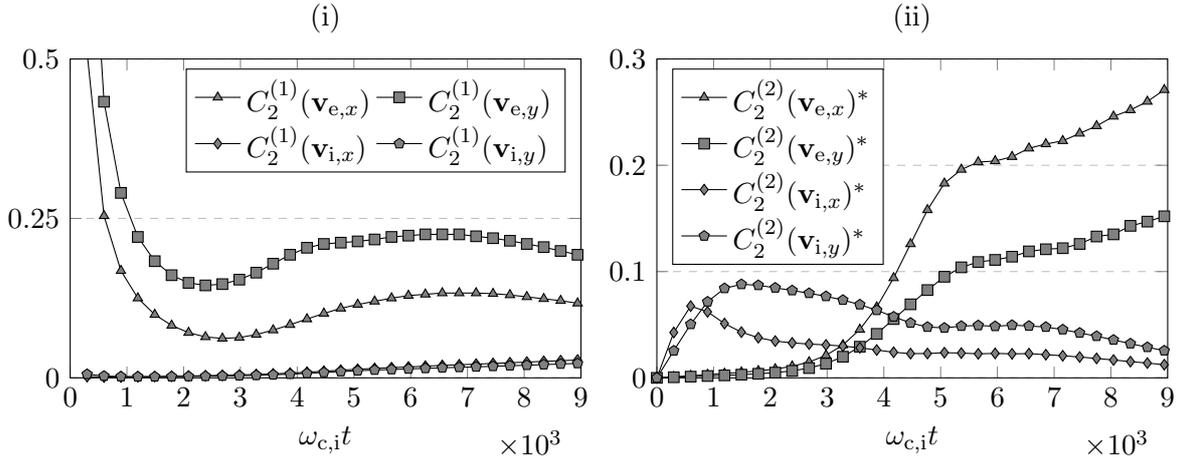

%% file: cold2.tex
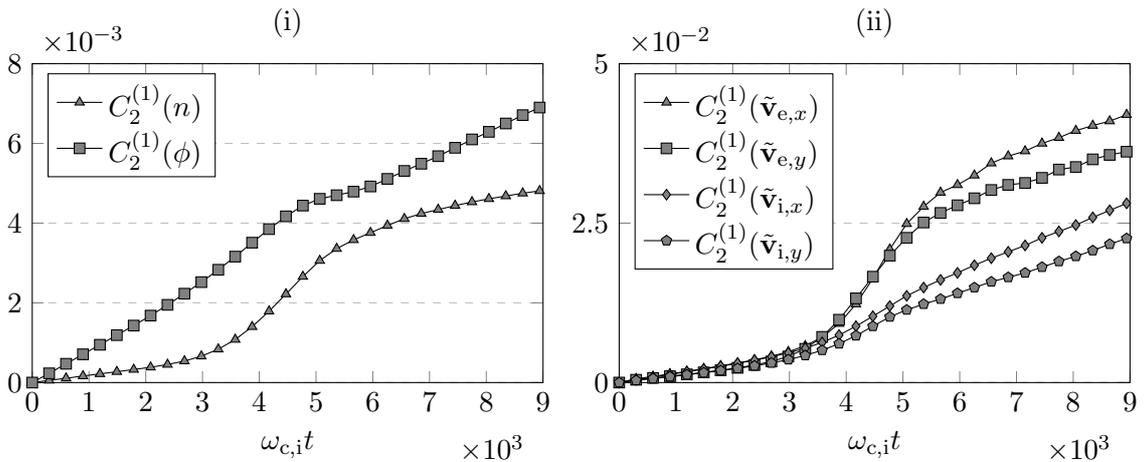
\begin{figure}[H]
\centering
\begin{tikzpicture}
\begin{groupplot}[
group style={rows=1, columns=2, vertical sep=2.0cm},
width=0.5\textwidth,
height=0.35\textwidth,
legend cell align={left},
cycle list name = my BandW,
]

\nextgroupplot[
title={(i)},
tick scale binop=\times,
xlabel={$\omega_{\text{c,i}} t$},
xmin=0, xmax=9000,
ymin=0, ymax=0.008,
xtick={0, 1000, 2000, 3000, 4000, 5000, 6000, 7000, 8000, 9000},
scaled x ticks={base 10:-3},
ytick={0, 0.002, 0.004, 0.006, 0.008},
/pgf/number format/.cd, 1000 sep={},
ymajorgrids=true,
grid style=dashed,
legend pos=north west,
] 
\addplot table [x={time}, y={n1e_n0}]{figures/sigma50.txt};
\addlegendentry{$C_2^{(1)}(n)$} 
\addplot table [x={time}, y={phi1_phi0}]{figures/sigma50.txt};
\addlegendentry{$C_2^{(1)}(\phi)$}

\nextgroupplot[
title={(ii)},
tick scale binop=\times,
xlabel={$\omega_{\text{c,i}} t$},
xmin=0, xmax=9000,
ymin=0, ymax=0.05,
xtick={0, 1000, 2000, 3000, 4000, 5000, 6000, 7000, 8000, 9000},
scaled x ticks={base 10:-3},
ytick={0, 0.025, 0.05},
/pgf/number format/.cd, 1000 sep={},
ymajorgrids=true,
grid style=dashed,
legend pos=north west,
] 
\addplot table [x={time}, y={vt1ex_vt0ex}]{figures/sigma50.txt};
\addlegendentry{$C_2^{(1)}(\tilde{\mathbf{v}}_{\text{e},x})$} 
\addplot table [x={time}, y={vt1ey_vt0ey}]{figures/sigma50.txt};
\addlegendentry{$C_2^{(1)}(\tilde{\mathbf{v}}_{\text{e},y})$}
\addplot table [x={time}, y={vt1ix_vt0ix}]{figures/sigma50.txt};
\addlegendentry{$C_2^{(1)}(\tilde{\mathbf{v}}_{\text{i},x})$} 
\addplot table [x={time}, y={vt1iy_vt0iy}]{figures/sigma50.txt};
\addlegendentry{$C_2^{(1)}(\tilde{\mathbf{v}}_{\text{i},y})$}

\end{groupplot}
\end{tikzpicture}
\caption{The operator $C$ applied on (i) the density and electric potential, and (ii) the corrected electron and ion velocity given by Eq.~\eqref{eq:estvel}. Both figures show the absolute values for the particular simulation parameters $\{\lambda=0, \tau_{\text{i}}=0, A=1, \sigma=50\}$. Note that in the quasi-neutral limit we have $n = n_{\text{e}} = n_{\text{i}}$. }
\label{fig:cold2}
\end{figure}

%% file: cold1.tex
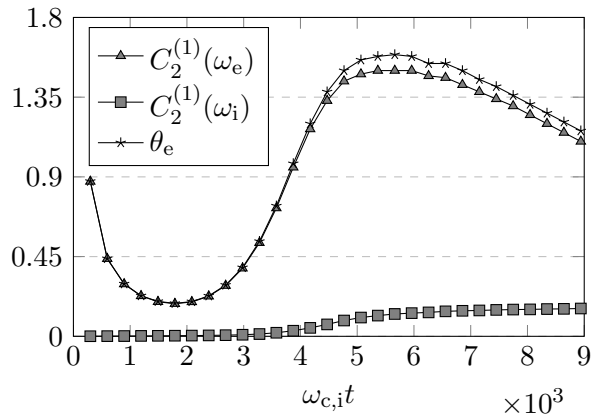
\begin{figure}[H]
\centering
\begin{tikzpicture}

\begin{groupplot}[
group style={rows=1, columns=2, vertical sep=2.0cm},
width=0.5\textwidth,
height=0.35\textwidth,
legend cell align={left},
cycle list name = my BandW,
]

\nextgroupplot[
tick scale binop=\times,
xlabel={$\omega_{\text{c,i}} t$},
xmin=0, xmax=9000,
ymin=0, ymax=1.8,
xtick={0, 1000, 2000, 3000, 4000, 5000, 6000, 7000, 8000, 9000},
scaled x ticks={base 10:-3},
ytick={0, 0.45, 0.9, 1.35, 1.8},
/pgf/number format/.cd, 1000 sep={},
ymajorgrids=true,
grid style=dashed,
legend pos=north west,
] 
\addplot table [x={time}, y={Oe1_O0}]{figures/sigma50_2.txt};
\addlegendentry{$C_2^{(1)}(\omega_{\text{e}})$} 
\addplot table [x={time}, y={Oi1_O0}]{figures/sigma50_2.txt};
\addlegendentry{$C_2^{(1)}(\omega_{\text{i}})$}
\addplot+[mark=star] table [x={time}, y={ODe1_O0}]{figures/sigma50_2.txt};
\addlegendentry{$\theta_{\text{e}}$}

\end{groupplot}
\end{tikzpicture}
\caption{The operator $C$ applied on the electron and ion vorticity along the magnetic field direction in comparison with the FLR parameter $\theta_{\text{e}}$ defined in the text. The figure show the absolute values for the particular simulation parameters $\{\lambda=0, \tau_{\text{i}}=0, A=1, \sigma=50\}$. }
\label{fig:cold1}
\end{figure}

%% file: hotiontable.tex
\begin{table}[H]
\centering
\ra{1.3}
\begin{tabular}{@{}lll@{}}
\toprule
Field & Time & Value \\
\midrule
$C_2^{(1)}(\phi)$ &
\(\displaystyle \sigma^{\frac{1}{2}} (1+\tau_{\text{i}})^{-\frac{1}{2}} \) &
\(\displaystyle \sigma^{-\frac{3}{2}} \tau_{\text{i}}^{\frac{3}{4}} \) \\[1mm]
$C_{2}^{(1)}(n)$ & 
\(\displaystyle \sigma^{\frac{1}{2}} (1+\tau_{\text{i}})^{-\frac{1}{2}} \) & 
\(\displaystyle \sigma^{-\frac{3}{2}} \tau_{\text{i}}^{\frac{3}{4}} \) \\[1mm]
$C_2^{(1)}(\mathbf{v}_{\text{e}})$, $C_2^{(1)}(\mathbf{\tilde{v}}_{\text{e}})$ & 
\(\displaystyle \sigma^{\frac{1}{2}} (1+\tau_{\text{i}})^{-1} \) &
\(\displaystyle \sigma^{-\frac{3}{2}} \) \\[1mm]
$C_2^{(1)}(\mathbf{v}_{\text{i}})$, $C_2^{(1)}(\mathbf{\tilde{v}}_{\text{i}})$ & 
\(\displaystyle \sigma^{\frac{1}{2}} (1+\tau_{\text{i}})^{-\frac{1}{2}} \) &
\(\displaystyle \sigma^{-\frac{3}{2}} \tau_{\text{i}}^{\frac{3}{4}} \) \\[1mm]
\bottomrule
\end{tabular}
\caption{Scaling relations for the operator $C$ applied on the fields in the regime where the correction to the ion velocity is dominated by the diamagnetic-drift velocity, i.e. at sufficiently large ion-temperature ratio $\tau_{\text{i}}$. 
} 
\label{tab:hotion}
\end{table}

%% file: hotevolution.tex
\begin{figure}[H]
\centering
\includegraphics[width=0.32\textwidth]{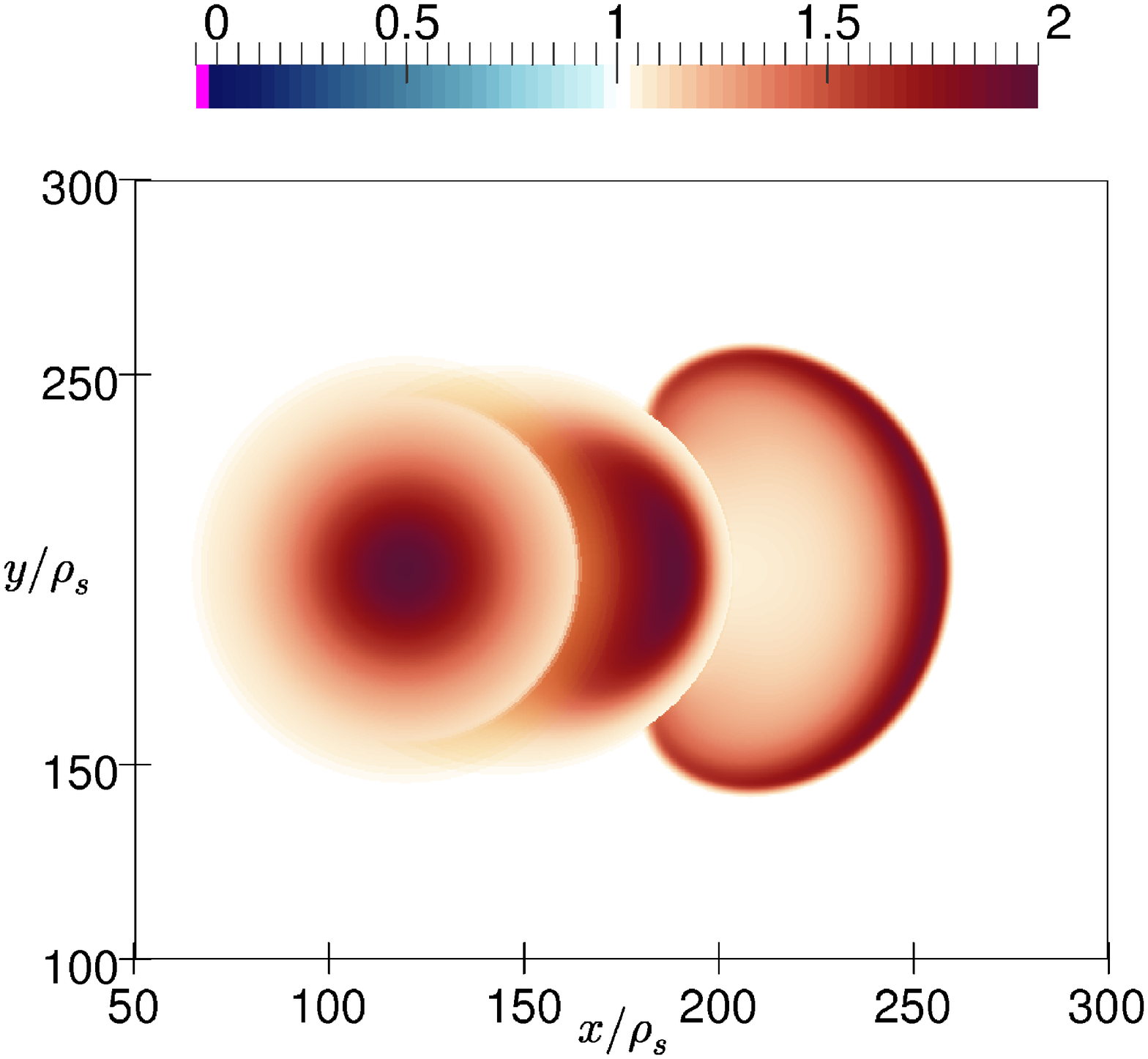}
\includegraphics[width=0.32\textwidth]{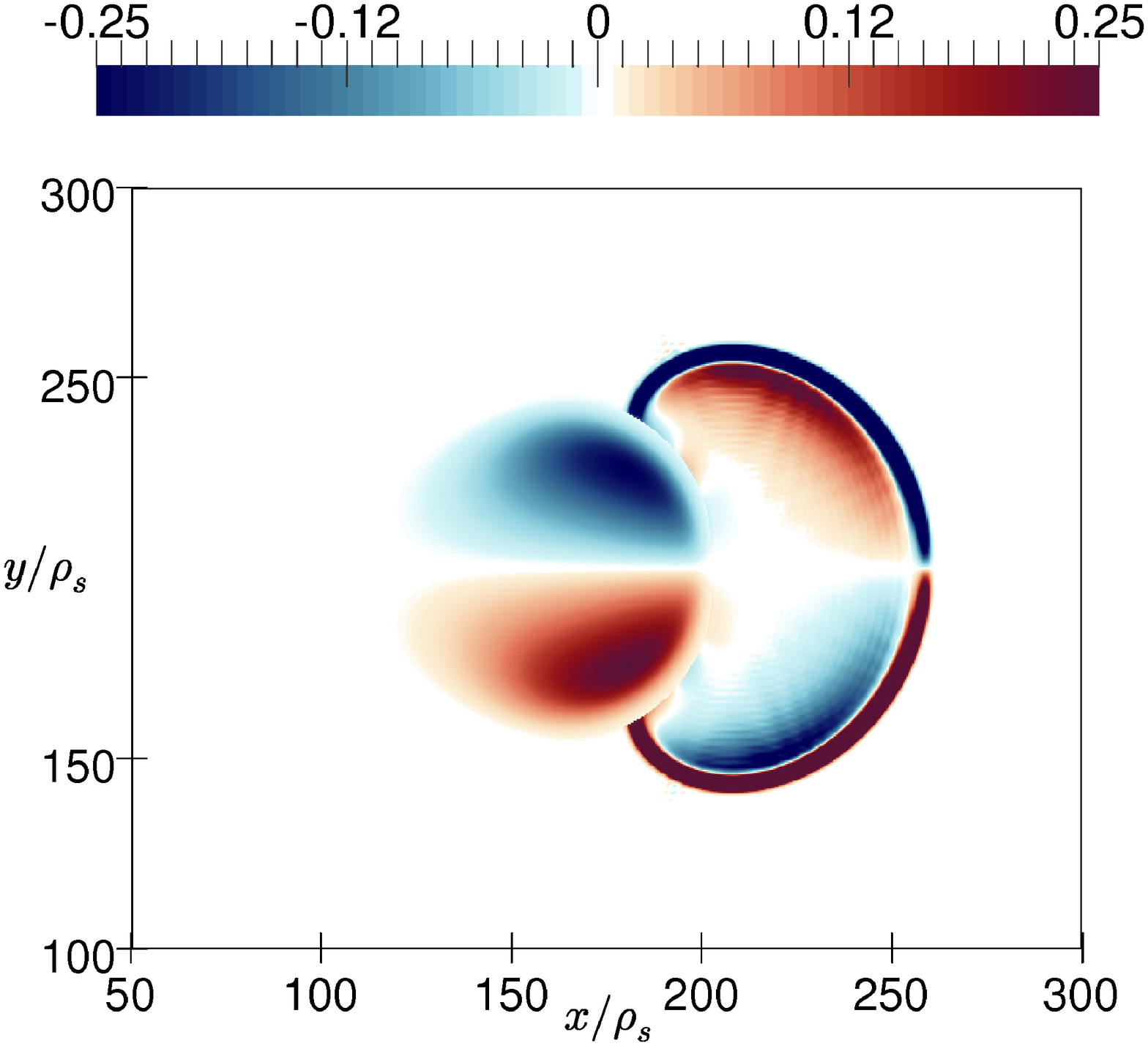}
\includegraphics[width=0.32\textwidth]{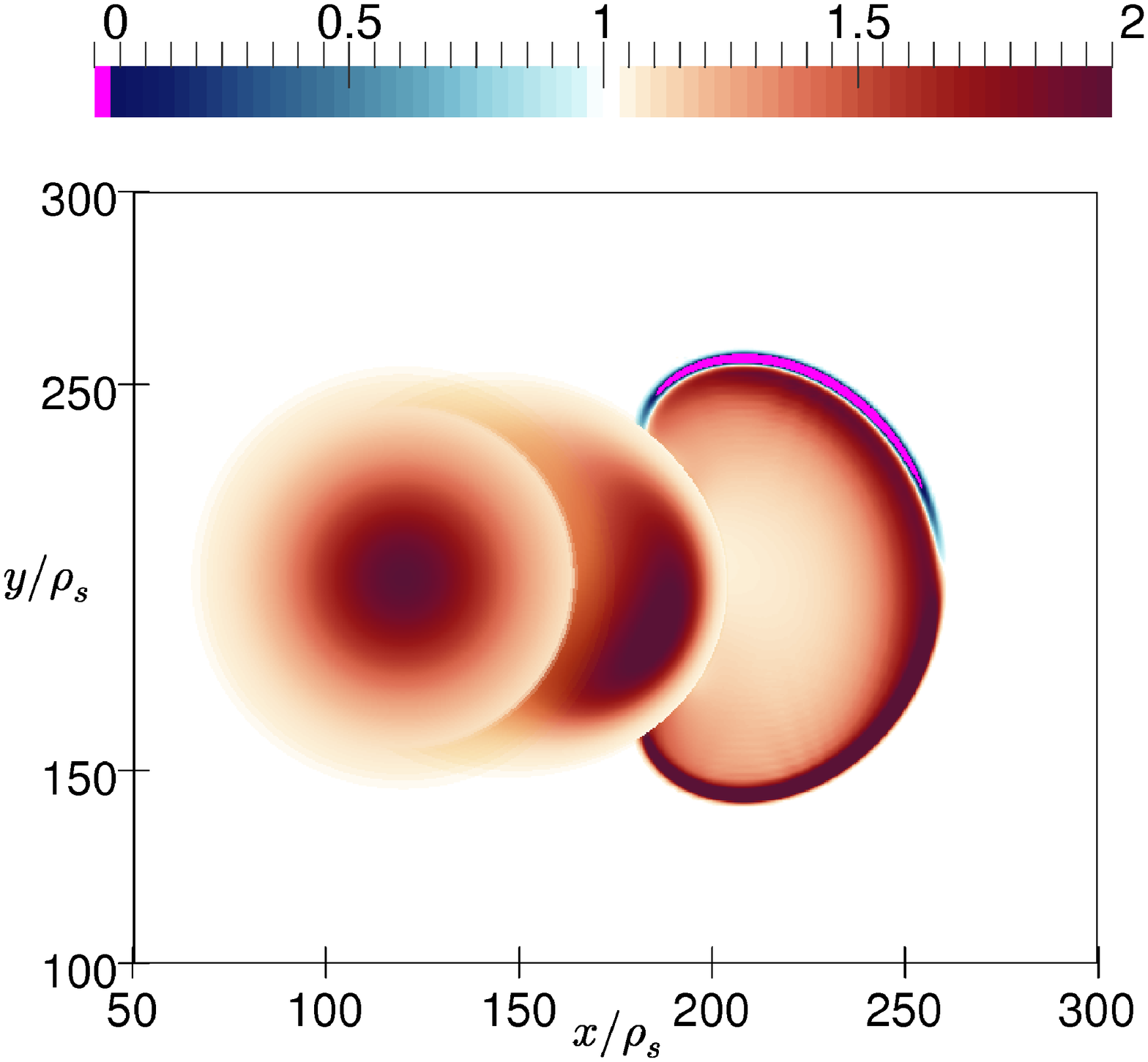}
\caption{The density field in the quasi-neutral limit $\lambda=0$ and non-zero ion temperature $\tau_{\text{i}} = 1$ for the parameters $A=1$ and $\sigma=20$. From left to right, the field $n^{(0)}$, $n^{(1)}$, and the \first-order approximation of the density $n \approx n^{(0)} + n^{(1)}$ are shown (overlapped) at three different times $\{0, 1330, 2261\}\omega_{\text{c,i}}$. The color scale for the \first-order correction has been cropped to show more of the structure.}
\label{fig:evolution1}
\end{figure}

%% file: tausigma.tex
\begin{figure}[H]
\centering
\begin{tikzpicture}
\begin{groupplot}[
group style={rows=1, columns=2, horizontal sep=2.3cm, vertical sep=0.0cm},
width=0.4\textwidth,
height=0.35\textwidth,
legend cell align={left},
cycle list name = my BandW,
]

\nextgroupplot[
title={(i)},
tick scale binop=\times,
xlabel={$\omega_{\text{c,i}} t$},
ylabel={$C^{(1)}_{\infty}(n)$},
xmin=0, xmax=6000,
ymin=0, ymax=1.0,
xtick={0, 1000, 2000, 3000, 4000, 5000, 6000},
scaled x ticks={base 10:-3},
ytick={0, 0.2, 0.4, 0.6, 0.8, 1.0},
/pgf/number format/.cd, 1000 sep={},
ymajorgrids=true,
grid style=dashed,
legend pos= outer north east,
legend style={font=\small},
] 
\addlegendentry{$\tau_{\text{i}} = 0$} 
\addlegendentry{$\tau_{\text{i}} = \frac{1}{4}$}
\addlegendentry{$\tau_{\text{i}} = \frac{1}{2}$}
\addlegendentry{$\tau_{\text{i}} = \frac{3}{4}$}
\addlegendentry{$\tau_{\text{i}} = 1$}
\addlegendentry{$\tau_{\text{i}} = 2$}
\addlegendentry{$\tau_{\text{i}} = 3$}
\addlegendentry{$\tau_{\text{i}} = 4$}

\addplot+[restrict x to domain=0:6000] table [x={time0}, y={n1e_n0_inf_0}]{figures/tauninf.txt};
\addplot+[restrict x to domain=0:5600] table [x={time}, y={n1e_n0_inf_025}]{figures/tauninf.txt};
\addplot+[restrict x to domain=0:5000] table [x={time}, y={n1e_n0_inf_05}]{figures/tauninf.txt};
\addplot+[restrict x to domain=0:4600] table [x={time}, y={n1e_n0_inf_075}]{figures/tauninf.txt};
\addplot+[restrict x to domain=0:3800] table [x={time}, y={n1e_n0_inf_1}]{figures/tauninf.txt};
\addplot+[restrict x to domain=0:2800] table [x={time}, y={n1e_n0_inf_2}]{figures/tauninf.txt};
\addplot+[restrict x to domain=0:2400] table [x={time}, y={n1e_n0_inf_3}]{figures/tauninf.txt};
\addplot+[restrict x to domain=0:2000] table [x={time}, y={n1e_n0_inf_4}]{figures/tauninf.txt};

\nextgroupplot[
title={(ii)},
tick scale binop=\times,
xlabel={$\omega_{\text{c,i}} t$},
xmin=0, xmax=4000,
ymin=0, ymax=1.0,
xtick={0, 1000, 2000, 3000, 4000},
scaled x ticks={base 10:-3},
yticklabels=\empty,
/pgf/number format/.cd, 1000 sep={},
ymajorgrids=true,
grid style=dashed,
legend pos= outer north east,
legend style={font=\small},
] 
\addlegendentry{$\sigma = 10$} 
\addlegendentry{$\sigma = 20$}
\addlegendentry{$\sigma = 30$}
\addlegendentry{$\sigma = 40$}
\addlegendentry{$\sigma = 50$}
\addlegendentry{$\sigma = 60$}
\addlegendentry{$\sigma = 70$}

\addplot+[restrict x to domain=0:1197] table [x={time}, y={n1e_n0_inf_10}]{figures/tausigma.txt};
\addplot+[restrict x to domain=0:2261] table [x={time}, y={n1e_n0_inf_20}]{figures/tausigma.txt};
\addplot+[restrict x to domain=0:2926] table [x={time}, y={n1e_n0_inf_30}]{figures/tausigma.txt};
\addplot+[restrict x to domain=0:3325] table [x={time}, y={n1e_n0_inf_40}]{figures/tausigma.txt};
\addplot table [x={time}, y={n1e_n0_inf_50}]{figures/tausigma.txt};
\addplot table [x={time}, y={n1e_n0_inf_60}]{figures/tausigma.txt};
\addplot table [x={time}, y={n1e_n0_inf_70}]{figures/tausigma.txt};

\end{groupplot}
\end{tikzpicture}
\caption{The coefficient $C^{(1)}_{\infty}(n)$ in the quasi-neutral limit $\lambda=0$ with $A=1$ for (i) fixed width $\sigma=40$ at various ion-temperature ratios $\tau_{\text{i}}$, and (ii) for fixed $\tau_{\text{i}}=1$ at different values of $\sigma$. For some domain of the ion-temperature ratio and width the \first-order corrections will be larger than the \zeroth-order density after some point in time.}
\label{fig:nInf}
\end{figure}
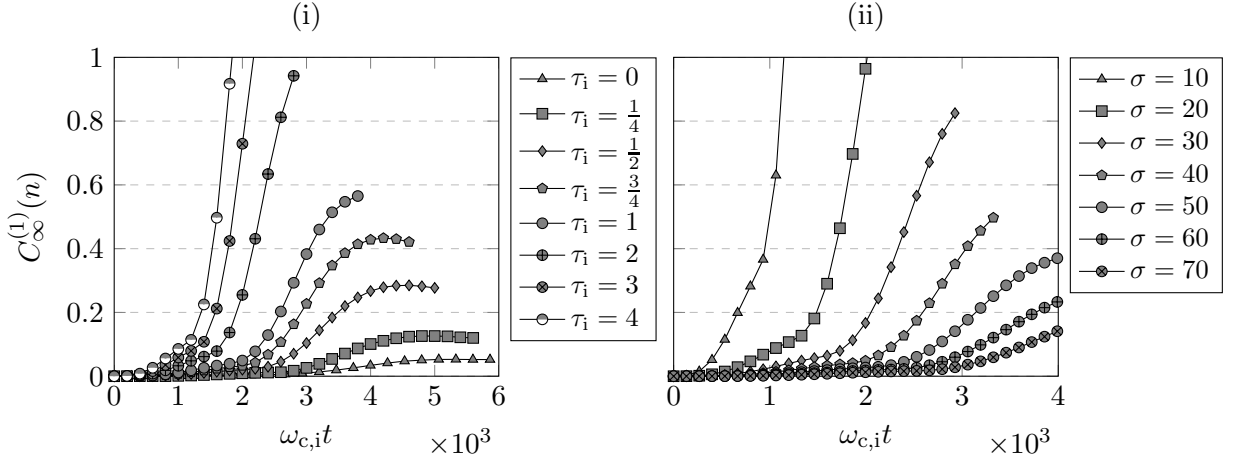

%% file: hot0.tex
\begin{figure}[H]
\centering
\begin{tikzpicture}
\begin{groupplot}[
group style={rows=1, columns=2, vertical sep=2.0cm},
width=0.5\textwidth,
height=0.35\textwidth,
legend cell align={left},
cycle list name = my BandW,
]

\nextgroupplot[
title={(i)},
tick scale binop=\times,
xlabel={$\omega_{\text{c,i}} t$},
xmin=0, xmax=5400,
ymin=0, ymax=0.04,
xtick={0, 1000, 2000, 3000, 4000, 5000, 6000},
scaled x ticks={base 10:-3},
ytick={0, 0.01, 0.02, 0.03, 0.04},
/pgf/number format/.cd, 1000 sep={},
ymajorgrids=true,
grid style=dashed,
legend pos=north west,
] 
\addplot table [x={time}, y={n1e_n0}]{figures/tauv.txt};
\addlegendentry{$C_2^{(1)}(n)$} 
\addplot table [x={time}, y={phi1_phi0}]{figures/tauv.txt};
\addlegendentry{$C_2^{(1)}(\phi)$}

\nextgroupplot[
title={(ii)},
tick scale binop=\times,
xlabel={$\omega_{\text{c,i}} t$},
xmin=0, xmax=5400,
ymin=0, ymax=0.4,
xtick={0, 1000, 2000, 3000, 4000, 5000, 6000},
scaled x ticks={base 10:-3},
ytick={0, 0.1, 0.2, 0.3, 0.4},
/pgf/number format/.cd, 1000 sep={},
ymajorgrids=true,
grid style=dashed,
legend pos=north west,
] 
\addplot table [x={time}, y={vt1ex_vt0ex}]{figures/tauv.txt};
\addlegendentry{$C_2^{(1)}(\tilde{\mathbf{v}}_{\text{e},x})$} 
\addplot table [x={time}, y={vt1ey_vt0ey}]{figures/tauv.txt};
\addlegendentry{$C_2^{(1)}(\tilde{\mathbf{v}}_{\text{e},y})$}
\addplot table [x={time}, y={vt1ix_vt0ix}]{figures/tauv.txt};
\addlegendentry{$C_2^{(1)}(\tilde{\mathbf{v}}_{\text{i},x})$} 
\addplot table [x={time}, y={vt1iy_vt0iy}]{figures/tauv.txt};
\addlegendentry{$C_2^{(1)}(\tilde{\mathbf{v}}_{\text{i},y})$}

\end{groupplot}
\end{tikzpicture}
\caption{The operator $C$ applied on (i) the density and electric potential, and (ii) the corrected electron and ion velocity given by Eq.~\eqref{eq:estvel}. Both figures show the absolute values for the particular simulation parameters $\{\lambda=0, \tau_{\text{i}}=\frac{1}{2}, A=1, \sigma=40\}$. Note that in the quasi-neutral limit we have $n = n_{\text{e}} = n_{\text{i}}$. }
\label{fig:hot0}
\end{figure}
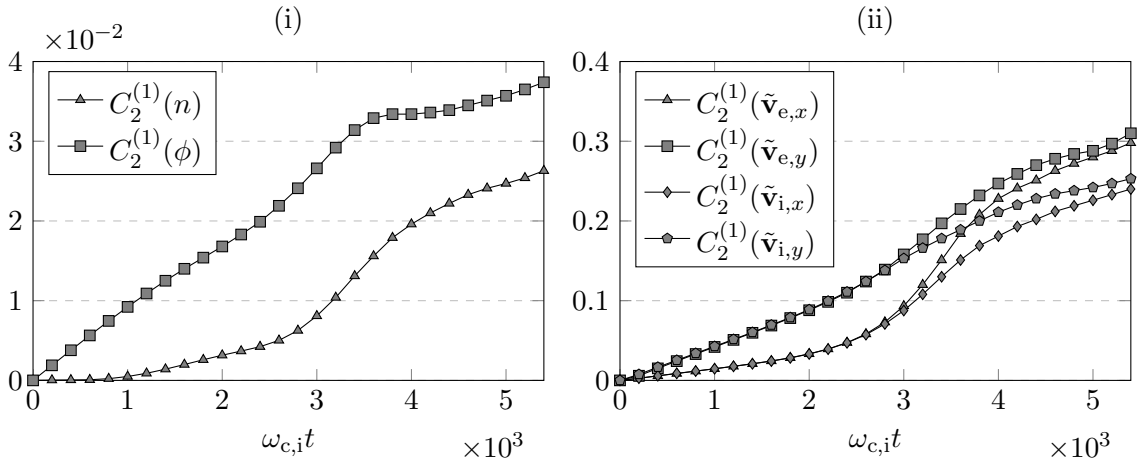

%% file: lambdatau0.tex
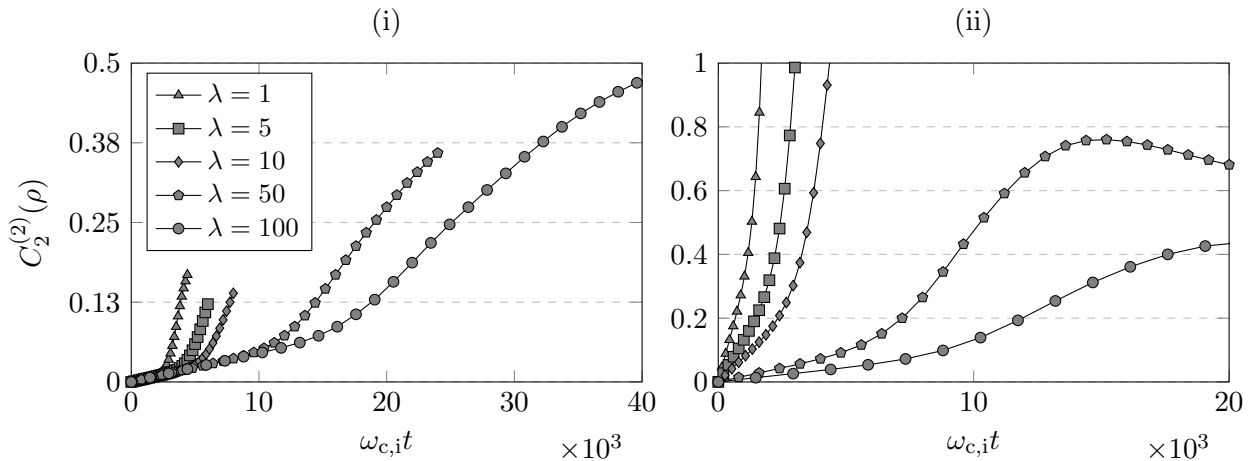
\begin{figure}[H]
\centering
\begin{tikzpicture}
\begin{groupplot}[
group style={rows=1, columns=2, vertical sep=2.0cm},
width=0.5\textwidth,
height=0.35\textwidth,
legend cell align={left},
cycle list name = my BandW,
]

\nextgroupplot[
title={(i)},
tick scale binop=\times,
xlabel={$\omega_{\text{c,i}} t$},
ylabel={$C^{(2)}_{2}(\rho)$},
xmin=0, xmax=40000,
ymin=0, ymax=0.5,
xtick={0, 10000, 20000, 30000, 40000},
scaled x ticks={base 10:-3},
ytick={0, 0.125, 0.250, 0.375, 0.5},
/pgf/number format/.cd, 1000 sep={},
ymajorgrids=true,
grid style=dashed,
legend pos=north west,
legend style={font=\small},
] 
\addlegendentry{$\lambda = 1$}
\addlegendentry{$\lambda = 5$}
\addlegendentry{$\lambda = 10$}
\addlegendentry{$\lambda = 50$}
\addlegendentry{$\lambda = 100$}

\addplot table [x={time1}, y={rho2_rho1_1}]{figures/lambdatau0.txt};
\addplot table [x={time5}, y={rho2_rho1_5}]{figures/lambdatau0.txt};
\addplot table [x={time10}, y={rho2_rho1_10}]{figures/lambdatau0.txt};
\addplot table [x={time50}, y={rho2_rho1_50}]{figures/lambdatau0.txt};
\addplot table [x={time100}, y={rho2_rho1_100}]{figures/lambdatau0.txt};

\nextgroupplot[
title={(ii)},
tick scale binop=\times,
xlabel={$\omega_{\text{c,i}} t$},
xmin=0, xmax=20000,
ymin=0, ymax=1.0,
xtick={0, 10000, 20000},
scaled x ticks={base 10:-3},
ytick={0, 0.2, 0.4, 0.6, 0.8, 1.0},
/pgf/number format/.cd, 1000 sep={},
ymajorgrids=true,
grid style=dashed,
] 

\addplot+[restrict x to domain=0:3370] table [x={time_1}, y={rho2_rho1_1}]{figures/lambdatau1.txt};
\addplot table [x={time_5}, y={rho2_rho1_5}]{figures/lambdatau1.txt};
\addplot table [x={time_10}, y={rho2_rho1_10}]{figures/lambdatau1.txt};
\addplot+[restrict x to domain=0:20000] table [x={time_50}, y={rho2_rho1_50}]{figures/lambdatau1.txt};
\addplot+[restrict x to domain=0:24931] table [x={time_100}, y={rho2_rho1_100}]{figures/lambdatau1.txt};

\end{groupplot}
\end{tikzpicture}
\caption{The $C_{2}$ operator applied on the charge density $\rho$ for (i) fixed $\{\tau_{\text{i}}=0, A=1, \sigma=20\}$ with varying lambda $\lambda$, and (ii) fixed $\{\tau_{\text{i}}=1, A=1, \sigma=20\}$ with varying lambda $\lambda$. The legend applies for both cases.}
\label{fig:chargedensity}
\end{figure}